\documentclass[final,3p,times,sort&compress]{elsarticle}

\usepackage{hyperref}
\hypersetup{colorlinks = false,linkcolor = blue,anchorcolor =red,citecolor = blue,filecolor = red,urlcolor = red, pdfauthor=author}
\usepackage{graphicx}
\graphicspath{{figures/}}
\usepackage{xcolor}
\usepackage{upgreek}
\usepackage{tikz}
\usetikzlibrary{calc,arrows,shadows,shapes,positioning}
\usepackage{siunitx}
\usepackage{miller}

\journal{Materials Characterization} 
\begin{document}

\begin{frontmatter}

\title{Observation of Bulk Plasticity in a Polycrystalline Titanium Alloy by Diffraction Contrast Tomography and Topotomography}

\address[label1]{Materials Science and Engineering, University of Illinois at Urbana-Champaign, IL 61801, USA}
\address[label2]{University of Lyon I, MATEIS INSA Lyon, CNRS UMR 5510, 69621 Villeurbanne, France}
\address[label3]{ESRF, The European Synchrotron, Grenoble 38043, France}
\address[label4]{Materials Science and Technology Division, The US Naval Research Laboratory, Washington, D.C. 20375, USA}
\address[label5]{Materials Department, University of California, Santa Barbara, CA 93106, USA}
\address[label6]{Institut PPRIME, Universit\'e de Poitiers, CNRS, ENSMA, UPR 3346, 86962 Chasseneuil Cedex, France}
\address[label7]{MINES ParisTech, PSL Research University, Centre des Matériaux, CNRS UMR 7633, BP 87, 91003, Evry, France}

\author[label1]{J. C. Stinville}

\author[label2,label3]{W. Ludwig}

\author[label4]{P. G. Callahan}

\author[label5]{M. P. Echlin}

\author[label6]{V. Valle}

\author[label5]{T. M. Pollock}

\author[label7]{H. Proudhon \corref{cor1}}
\cortext[cor1]{corresponding author}
\ead{henry.proudhon@mines-paristech.fr}

\begin{abstract}
%The investigation during deformation of the location of the slip events as a function of the 3D microstructure of polycrystalline metallic materials is critical to the understanding of the effect of microstructure on mechanical properties. 

The mechanical properties of polycrystalline metals are governed by the interaction of defects that are generated by deformation within the 3D microstructure. In materials that deform by slip, the plasticity is usually highly heterogeneous within the microstructure. Many experimental tools can be used to observe the results of slip events at the free surface of a sample; however, there are only a few methods for imaging these events in the bulk. In this article, the imaging of bulk slip events within the 3D microstructure are enabled by the combined use of X-ray diffraction contrast tomography and topotomography. Correlative measurements between high-resolution digital image correlation, X-ray diffraction contrast tomography, topotomography and phase contrast tomography are performed during deformation of Ti-7Al to investigate the sensitivity of the X-ray topotomography method for the observation of slip events in the bulk. Much larger neighborhoods of grains were able to be mapped than in previous studies, enabling quantitative measurements of slip transmission. Significant differences were observed between surface and bulk grains, indicating the need for 3D observations of plasticity to better understand deformation in polycrystalline materials.
%Compared to previous studies, much larger neighborhoods of grains could be mapped that allowed quantitative measurements of slip transmission and showed significant differences between bulk and surface grains. %Furthermore, new opportunities for using X-ray topotomography to investigate bulk slip events and slip transmission are demonstrated.  
\end{abstract}


\end{frontmatter}

\section{Introduction \& Background}

% Systematic / Quantitative / Statistical Topotomo Framework
    % \subsection{Diffraction Contrast Tomography}
    % \subsection{Digital Image Correlation}
    % \subsection{Topotomography}

% Application to \subsection{Ti-7Al Mechanical Behavior / Plasticity Mechanisms}
    %Intro To Titanium alloys 
    %Intro To Ti-7Al 
    %>4% Al - > alpha precips 
    %Localization in Ti-7Al affect of precipitats 
    %Slip Localization in materials
    %Plasticity \& localization community 

During plastic deformation, metallic materials may experience localized slip on glide planes over numerous inter-atomic distances \cite{Mitchell1993}. As a consequence of the formation of slip bands within a material, steps form where the slip bands intersect the surface of the sample and can be observed using optical \cite{Lombros2016}, confocal \cite{LIU2019260}, scanning electron \cite{Stinville2015expmech}, and atomic force microscopy \cite{BONNEVILLE200887,AUBERT20169}. Information about the deformation processes that occur during monotonic and cyclic loading is obtained by analyzing the slip traces and their characteristics at the surface in relation to the microstructure. Such information has provided a deep understanding of deformation processes that occur in the plastic regime. For instance, the analysis of surface slip traces in the micro-plastic regime (plasticity prior to macroscopic yield) \cite{MAA2018338} has provided insights on dislocation interactions, avalanche processes and incipient slip and their effect on mechanical properties \cite{Fan2021}. The development of high-resolution digital image correlation (HR-DIC) inside the scanning electron microscope (SEM) provided opportunities to statistically investigate incipient slip in metallic materials \cite{Chen2018}, and to relate the intensity of surface slip events to fatigue properties \cite{STINVILLE2020172}. The analysis of slip traces by SEM, focused ion beam SEMs (FIB-SEMs) and/or atomic force microscopy for specimens deformed cyclically in the macroscopic plastic regime has led to the identification of the micro-scale characteristics of fatigue structures such as persistent slip bands or deformation bands \cite{MAN20101625,Mughrabi2009,Ho2015}. These analyses have provided new insights into fatigue crack nucleation mechanisms, and in the high strain regime further understanding of the strain hardening behavior \cite{WELSCH2016188,WIECZOREK2016320}. The experimental observation/analysis of slip traces is also extensively used to study the transmission of slip across interfaces \cite{Kacher2014}. Most transmission models are based on geometric or orientation-driven relationships with incorporated mechanical loading models \cite{Bayerschen2015}. For example, the m' factor \cite{Luster1995} successfully describes surface slip transmission in titanium alloys \cite{BIELER2014212,HEMERY2018277}. These models have been validated using 2D experimental surface observations of slip transmission, which may be influenced by free surface effects. Slip transmission may significantly differ in the bulk and has not been investigated until now.

While there exist many experimental tools to observe surface or near sub-surface slip events, there are few methods for gathering this information in the bulk. Techniques where thin foils or lamellae are imaged in a electron transmission configuration, including the transmission electron microscope (TEM) \cite{SURI19991019} and transmission SEM (known as t-SEM or SEM STEM) \cite{STINVILLE2019152} can probe slip events in a sample extracted from the interior of a sample. However, the sample lamellae are extremely small and thin (typically \SI{10}{\micro\meter} $\times$ \SI{10}{\micro\meter} $\times$ \SI{100}{\nano\meter}), which prevents the statistical analysis of slip over large regions containing many grains with varying orientations, or during complex loading conditions.

Over the past two decades, synchrotron X-ray techniques have been developed that enable the imaging of grains \cite{Larson:2002,Poulsen2012,Suter2006,Ludwig2009,Bernier2020} and sub-grain structure \cite{Ludwig2007,Vigano2016,Nygren2020}, stress states within grains \cite{Henningsson2020,Reischig_COSS_2020,Shen2020, Miller2020}, and recently dislocations or slip events within grains using dark field X-ray microscopy and topotomography \cite{yildirim_2020,Proudhon2018}. The recent increase in spatial resolution and detector sensitivity due to upgrades to synchrotron light source flux (ESRF-EBS upgrade \cite{Cho2020,ESRF-EBS-upgrade}) and advances in detector technology (i.e. single photon counting pixel detectors and back-illuminated  sCMOS cameras) provide the opportunity to directly image plastic deformation events throughout the interior of mm$^3$-scaled samples. The combination of X-ray diffraction contrast tomography (DCT), topotomography and phase contrast tomography (PCT) has been particularly effective for imaging slip events in the bulk polycrystalline structure, in situ during deformation and then relating these events to the 3D grain structure \cite{Proudhon2018}. However, the mechanisms giving rise to image contrast associated with slip events in the X-ray topographs are not well understood and the sensitivity of the technique for capturing all slip events was not demonstrated. 

%Be sure to talk about Wolfgang's previous measurements of angular resolution resolvable in topotomo \cite{XYZ} in the context of DIC measurements. 

In this research, quantitative correlative measurements between HR-DIC, X-ray DCT and topotomography are performed on a Ti-7Al titanium alloy in order to investigate the sensitivity of the X-ray topotomography method to probe slip events both at the surface and in the bulk. The opportunities provided by the X-ray topotomography method to investigate plastic localization and transmission in metallic materials are demonstrated and discussed.

\section{Methods}

%A custom sample was prepared such that topotomography data could be collected from a single sample loaded in tension, while also being compatible with DCT, SEM-DIC, and EBSD imaging modes.

    \subsection{Ti-7Al Structure and Sample Preparation} \label{sec:MethodsSample}
    
Ti-7Al was selected as a model material in the present work due to its relatively large equiaxed grain structure and extensive study in the literature \cite{CHATTERJEE201635,Lienert2009,Venkataraman2017}. This material was extruded and well-annealed to produce large ($\sim$\SI{100}{\micro\meter}), fully recrystallized grains with minimal intragranular substructure \cite{CHATTERJEE201635}. The Ti-7Al material contains $\alpha_2-\mathrm{Ti}_\mathrm{3}\mathrm{Al}$ precipitates that are roughly 2-3 nm in size within the $\alpha$ hexagonal close packed matrix phase. More details on the Ti-7Al alloy structure can be found elsewhere \cite{Venkataraman2017}.  

The specimen geometry used for correlative measurements is presented in \autoref{DIC_ts3}(a) and was prepared by wire electrical discharge machining to minimize residual stresses and plastic deformation from machining. The gauge section of the sample is rectangular with a cross sectional area of \SI{0.6}{\milli\meter} x \SI{0.6}{\milli\meter}. 
The flat sections of the gauge surface orthogonal to the loading direction were mechanically mirror polished using abrasive papers and diamond suspension, then chemo-mechanically polished using a suspension of \SI{0.04}{\micro\meter} colloidal silica particles. Prior to deformation, a speckle pattern was obtained by chemical etching by immersion in Kroll's reagent for \SI{20}{s}.

\subsection{High Resolution Digital Image Correlation and microstructure measurements}

Tensile tests were performed using a custom in situ $\pm$ 5000 N stage within a Thermo Fisher Scientific Versa3D microscope with a field-emission electron emitter on the flat dogbone-shaped specimens described in \autoref{sec:MethodsSample}. Tensile tests were interrupted at a macroscopic strain level near 0.8\% (just past the 0.2\% offset yield strength) and then HR-DIC measurements were performed. Macroscopic strain was measured in situ using both a strain gauge and fiducial markers located at both ends of the gauge length.     
    
\begin{figure}
    \centering
    \includegraphics[width=1\textwidth]{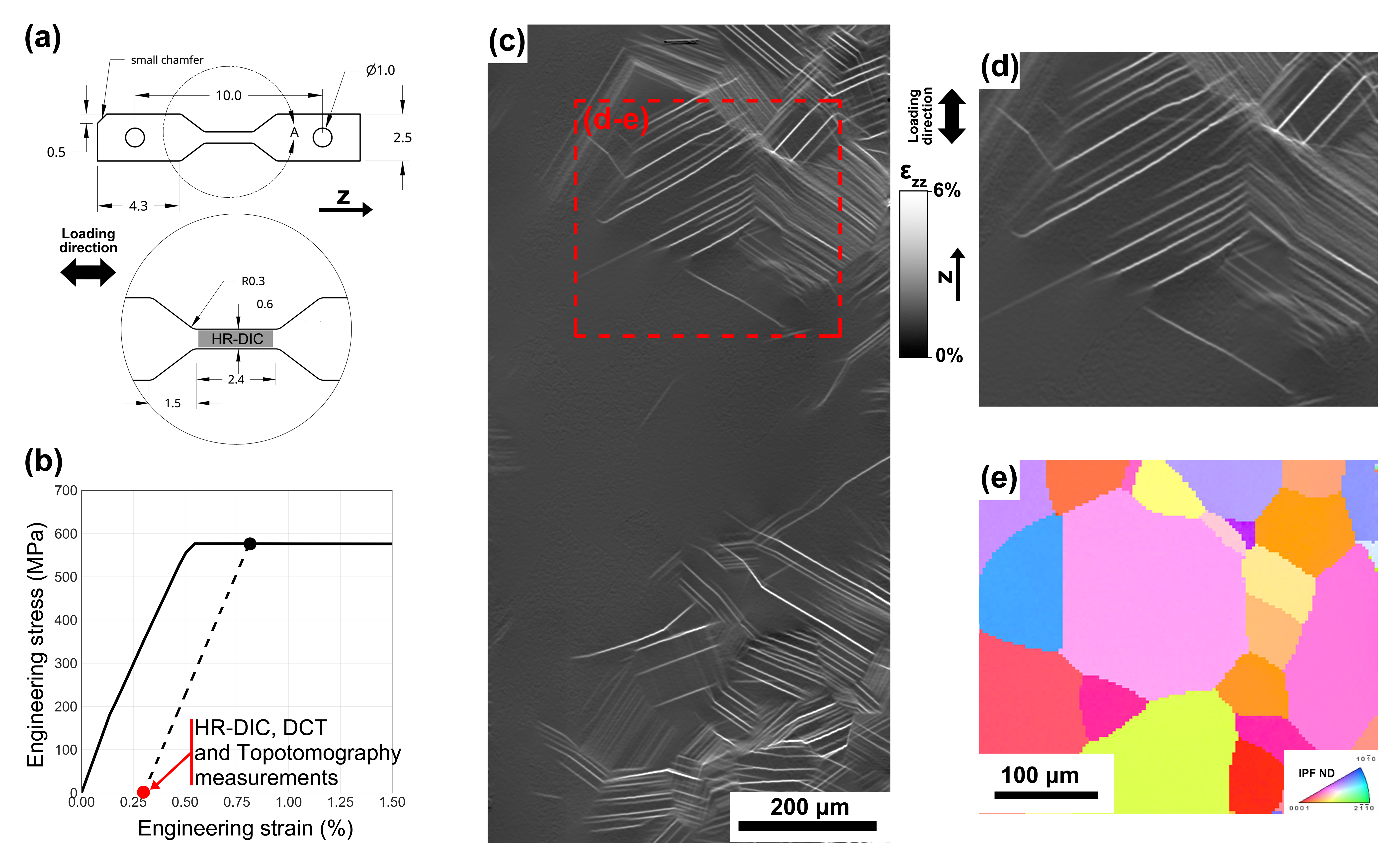}
    \caption{(a) Specimen geometry used for correlative in situ HR-DIC, DCT and topotomography measurements. (b) Engineering stress-strain curve of the Ti-7Al material. (c) HR-DIC $\epsilon_{zz}$ strain map obtained at the surface of the specimen after about 0.8\% macroscopic deformation. (d-e) Reduced region of the $\epsilon_{zz}$ strain map with its associated grain structure obtained by electron backscatter diffraction measurement.}
    \label{DIC_ts3}
\end{figure}

SEM image sets were acquired before loading and while under load following the guidelines of Kammers and Daly \cite{Kammers_2013a,Kammers_2013b} and Stinville et al \cite{Stinville_2015a}. A National Instruments\textsuperscript{TM} scan controller and acquisition system (DAQ) was used to control electron beam scanning in the microscope. This custom beam scanner removes the SEM beam defects associated with some microscope scan generators \cite{Stinville_2015a,LENTHE201893}. Tiles of $8 \times 4$ images before and after deformation with an image overlap of 15\% were used. HR-DIC calculations were performed on these series of images and the results merged using a pixel resolution merging procedure, which is described in detail elsewhere \cite{Chen2018}. HR-DIC measurements were performed on the entire gauge section surface, an area of about \SI{2}{\milli\meter} $\times$ \SI{600}{\micro\meter}. Typical subset size values of $31 \times 31$ pixels (\SI{1044}{\nano\meter} $\times$ \SI{1044}{\nano\meter}) with a step size of 3 pixels (\SI{101}{\nano\meter}) were used for the DIC measurements. Digital image correlation was performed using the Heaviside-DIC method \cite{BOURDIN2018307,Valle2015}. The sample preparation, imaging conditions and Heaviside-DIC parameters enable the detection of slip events at the surface of the specimen with a discontinuous displacement resolution between 0.2 and 0.3 pixels (\SI{7}{\nano\meter} and \SI{10}{\nano\meter} respectively) \cite{BOURDIN2018307,Valle2015}. The strain map of the investigated specimen deformed at 0.27\% plastic deformation is displayed in \autoref{DIC_ts3}(c and d) for the entire gauge length and for a reduced region of interest (ROI), respectively. The axial loading direction is vertically oriented in all strain maps. Bands of concentrated strain identify the locations of slip events at the surface of the specimen. It is worth noting that enhanced plasticity occurred near the edges of the gage section of the specimen due to slight misalignment during mechanical loading.    

Microstructure characterization on the sample gauge surface was performed by electron backscatter diffraction (EBSD) measurements with an EDAX OIM-Hikari XM4 EBSD detector using step size of \SI{0.5}{\micro\meter}. Diffraction patterns were acquired using an accelerating voltage of \SI{30}{\kilo\volt}, a $4 \times 4$ binning mode and a beam current of \SI{0.2}{\nano\ampere}. EBSD maps were acquired before and after deformation. The data registration between the gauge surface EBSD and HR-DIC measurements is done by aligning control points using a polynomial distortion. About 50 pairs of control points were picked over the area of interest. The procedure is detailed elsewhere \cite{Charpagne2019}.    

\subsection{Diffraction Contrast Tomography} \label{sec:ts3_dct}

X-ray characterization was carried out at the ESRF materials science beam line ID11 in Grenoble, France. The beam energy was set to 38 keV with a relative bandwidth of $3\times10^{-3}$. The experimental setup includes a diffractometer designed for 3D diffraction experiments, including DCT and topotomography imaging modalities \cite{Lud01a,Ludwig:2007ab} (see Section~\ref{sec:ts3_tt_setup}).

The Ti-7Al specimen was mounted vertically on the rotation stage. The microstructure in the gauge length was first characterized by two partially overlapping DCT scans, each composed of 3600 equally spaced projections over \SI{360}{\degree}. Images were recorded with an exposure of \SI{0.25}{s} on a high-resolution detector composed of a scintillator placed 6.5 mm after the sample that is optically coupled by a $10\times$ objective to a 2048 $\times$ 2048 pixel ESRF Frelon camera, giving an effective pixel size of \SI{1.4}{\micro\meter}. The total scan time was 30 minutes.

%\begin{figure}
%\centering
%\begin{tikzpicture}
%\node[above] at (-4,0) %{\includegraphics[width=8cm]{figures/ts3_dct_merged.png}};
%\node at (-4,0) {(a)};
%\node[above] at (4,0.5) %{\includegraphics[width=8cm]{figures/ts3_grain_size.pdf}};
%\node at (4,0) {(b)};
%\end{tikzpicture}
%\caption{(a) Diffraction Contrast Tomography of the Ti-7Al sample %(random colors are used for the grains), the front face corresponds %very well with EBSD recorded before the synchrotron experiment. (b) %Grain size plot from the DCT measurement.}
%\label{fig:ts3_dct}
%\end{figure}
% - see \autoref{fig:ts3_dct}b

Each scan was reconstructed using the DCT software \cite{dct_code} to produce grain maps of the bulk region subsurface below the HR-DIC measurements that also contained the topotomography measurements. The transmission of the direct beam was used to reconstruct the exact shape of the illuminated region using a classical filtered back projection. The two DCT grain volumes were fused together into one dataset using Pymicro \cite{Pymicro} (including absorption data to produce a mask for the specimen geometry) after matching grains in the overlapping region and minimizing the correlation error between the two DCT volumes present in the overlapped region. Isotropic dilation was also carried out, but limited to the grains within the absorption mask, in order to fill all remaining void space within the DCT reconstruction. This resulted in a high fidelity grain map of the entire ROI of the sample (\SI{0.6}{\milli\meter} $\times$ \SI{0.6}{\milli\meter} $\times$ \SI{1.1}{\milli\meter}), containing 1055 grains. The average grain size in the reconstructed DCT volume, including surface grains, is \SI{68}{\micro\meter}. As a side note, besides the front face that was mirror polished for EBSD analysis, the other 3 lateral faces were left as received from the EDM machining. This resulted in a \SI{10}{\micro\meter} thick non-diffracting layer of material that is absent from the final reconstruction. 

\subsection{Topotomography setup} \label{sec:ts3_tt_setup}
The reconstructed grain map (see Section \ref{sec:ts3_dct}) was processed to select grains for topotomography inspection. After matching the EBSD data to the surface of the reconstructed DCT grain map, two surface grains (labeled 63 and 567 and visible on \autoref{fig:Transmission_1}) with surface slip traces identified by HR-DIC were selected from two different locations. More grains were selected based on the following criteria: (i) the grain is present in the neighborhood of the two selected surface grains (ii) at least one low index reflection can be aligned within the range of the sample goniometer. An automated script searched through all the indexed grains and selected 55 different grains for topotomography acquisition. For some of these grains, several reflections (up to 3) were available within the goniometer ranges and were also collected. In total, 75 topotomography scans were collected for two grain clusters in the gauge length of the sample.

\begin{figure}
    \centering
    \includegraphics[width=1\textwidth]{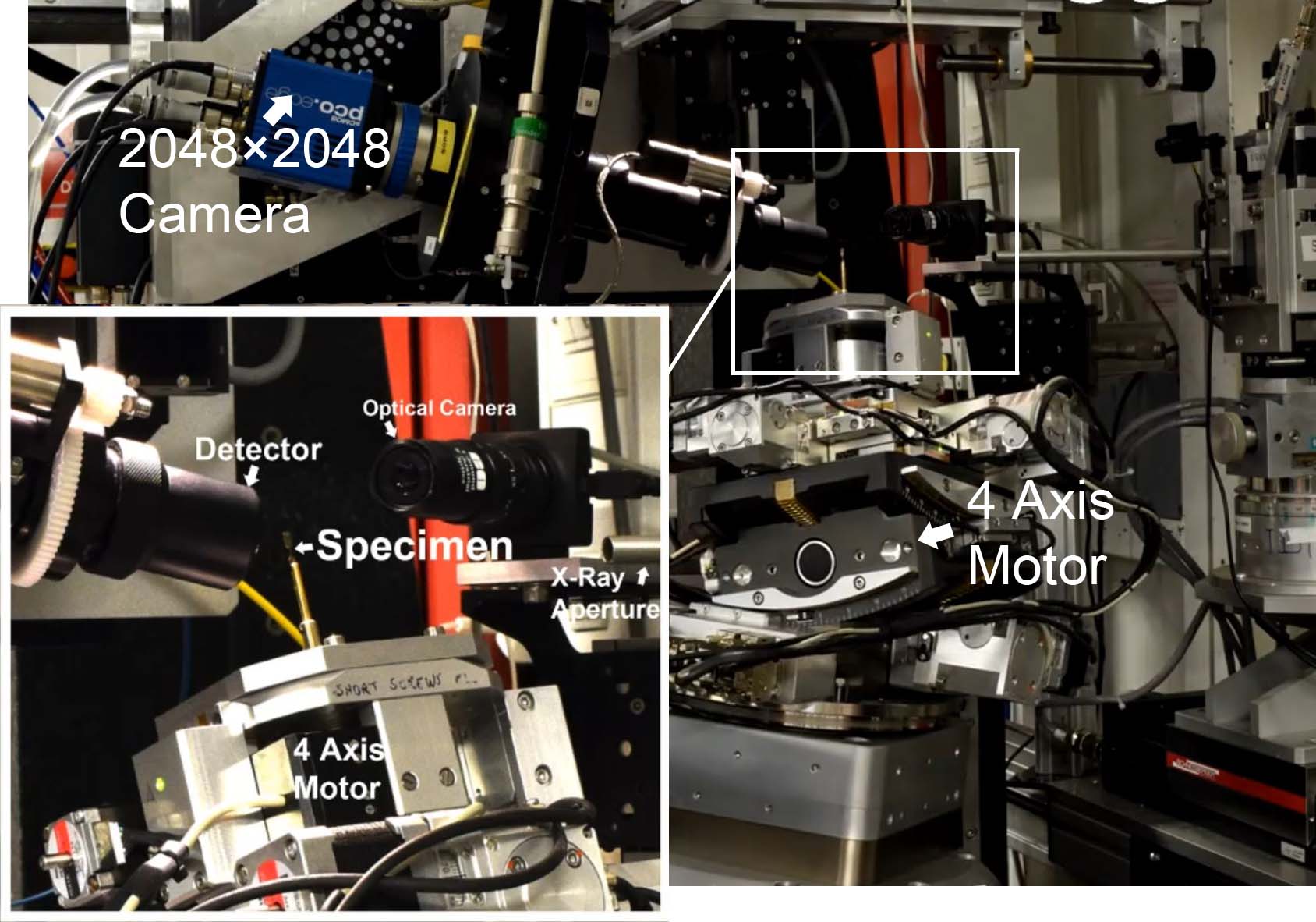}
    \caption{Experimental setup for topotomography measurement. The complete setup is rotated by the base tilt $\Theta$ while maintaining Bragg condition visibility for a specific grain throughout the entire \SI{360}{\degree} rotation around the tomographic rotation axis. A high resolution detector with a pixel size of \SI{0.7}{\micro\meter} is used to record series of integrated projection topographs every \SI{4}{\degree}. The reader is invited to consult the Supplemental Material on the experimental setup for topography measurement.}
    \label{fig:TopoTomoSetup}
\end{figure}

For topotomographic scan acquisition a particular scattering vector of a given grain is aligned parallel to the tomographic rotation axis \cite{Ludwig2007}. The tomographic rotation axis itself is inclined by the Bragg angle $\Theta$ (diffractometer base tilt) in order to maintain Bragg condition throughout the entire \SI{360}{\degree} rotation of the sample. To account for lattice rotations and dispersion effects within the grain, the base tilt is scanned over the width of the crystal reflection curve using an alternating, continuous scan motion with a step size of \SI{0.05} degrees. A second camera with a higher magnification was used to record these projection topographs with a pixel size of \SI{0.7}{\micro\meter}. In the current experiment, topotomography scans were comprised of 90 equispaced series of projection topographs (one every \SI{4}{\degree} of sample rotation) and require between 10 minutes to a couple of hours to collect, depending on the level of mosaicity (the $\Theta$ integration range must be increased for more deformed grains, a typical value was used for this sample with a range of \SI{1.6}{\degree} and 64 images). The topotomography experimental setup is presented in the Supplementary Material, including a high-speed video (20 $\times$ speed) of the sample goniometer stages moving during data collection. A snapshot of the video is provided in \autoref{fig:TopoTomoSetup} to describe the experimental setup. An example of a set of topographs is provided in Supplementary Material for a grain that is subject to plastic deformation. 

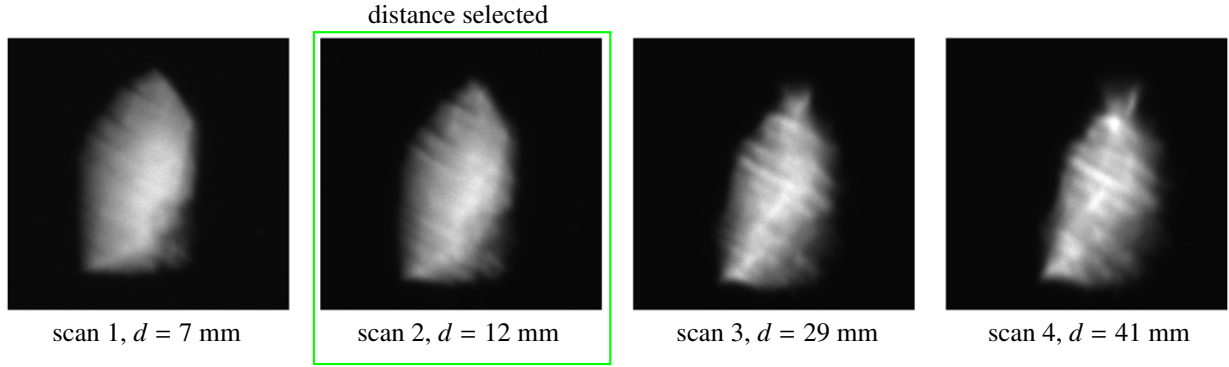
\begin{figure}
\begin{tikzpicture}
\node[above] {\includegraphics[width=\textwidth]{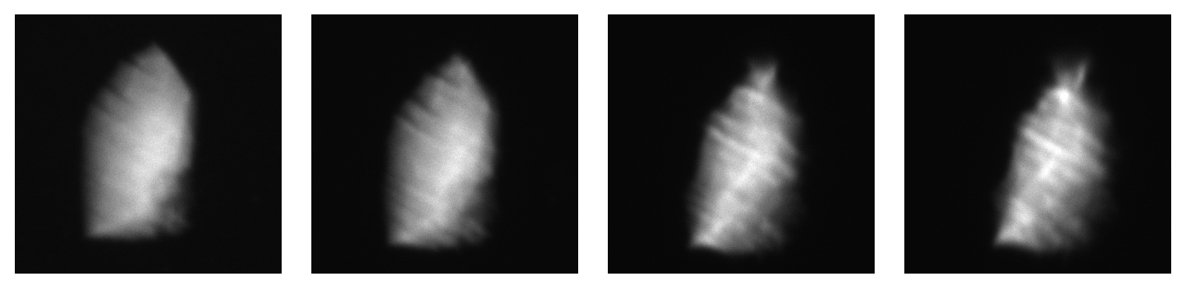}};
\node at (-6.2,0) {scan 1, $d=7$ mm};
\node at (-2.1,0) {scan 2, $d=12$ mm};
\node at (2.1,0) {scan 3, $d=29$ mm};
\node at (6.2,0) {scan 4, $d=41$ mm};
\draw[green,thick] (-4,-0.4) rectangle (-0.1,4.0);
\node[above] at (-2.1,4.0) {distance selected};
\end{tikzpicture}
\caption{Topographs of grain 567 extracted from the topotomography scans ($\omega$=\SI{44}{\degree}) taken at different detector distances. The second distance of \SI{12}{\milli\meter} was selected for our observations. A movie showing the evolution of contrast as a function of the detector distance is available as Supplementary Material.}
\label{fig:ts3_tt_nfdtx}
\end{figure}

X-ray topographs contain orientation contrast from the diffraction of a crystal. Defects present in the volume locally affect the Bragg condition and will give rise to contrasts evolving as a function of the propagation distance. To select the best detector distance to acquire topotomography scans, grain 567 was first aligned using reflection \hkl(1 1 -2 0) and integrated projection topographs were recorded while increasing the sample detector distance from \SI{7}{\milli\meter} to \SI{41}{\milli\meter} in steps of \SI{2}{\milli\meter}. The contrast related to slip events evolves from low contrast intensity at short propagation distance to high contrast intensity at large propagation distance, whereas the outline of the grain evolves from a nearly geometric projection into a distorted projection with ill defined boundaries at large distances, as shown in \autoref{fig:ts3_tt_nfdtx}. For the rest of the experiment the  detector distance was kept at \SI{12}{\milli\meter}, allowing enough space to rotate the tension specimen freely without risking a collision bewteen the goniometer stage and the detector. Note that in situ testing (not reported here) requires an optimized design of the load frame \cite{Proudhon2018} and of the detector head in order to maintain such short propagation distances. Also note that in for future experiments, acquisitions with different propagation distances provide additional constraints which can improve the convergence of the iterative optimization algorithms used for reconstructing the local orientation field inside a grain \cite{Vigano_COSS_2020}.

%\ref{Supsec:TopotomographySetup}

\subsection{Detecting slip system activation}
\begin{figure}[htb]
\centering
\begin{tikzpicture}
\node[above] {\includegraphics[width=0.8\textwidth]{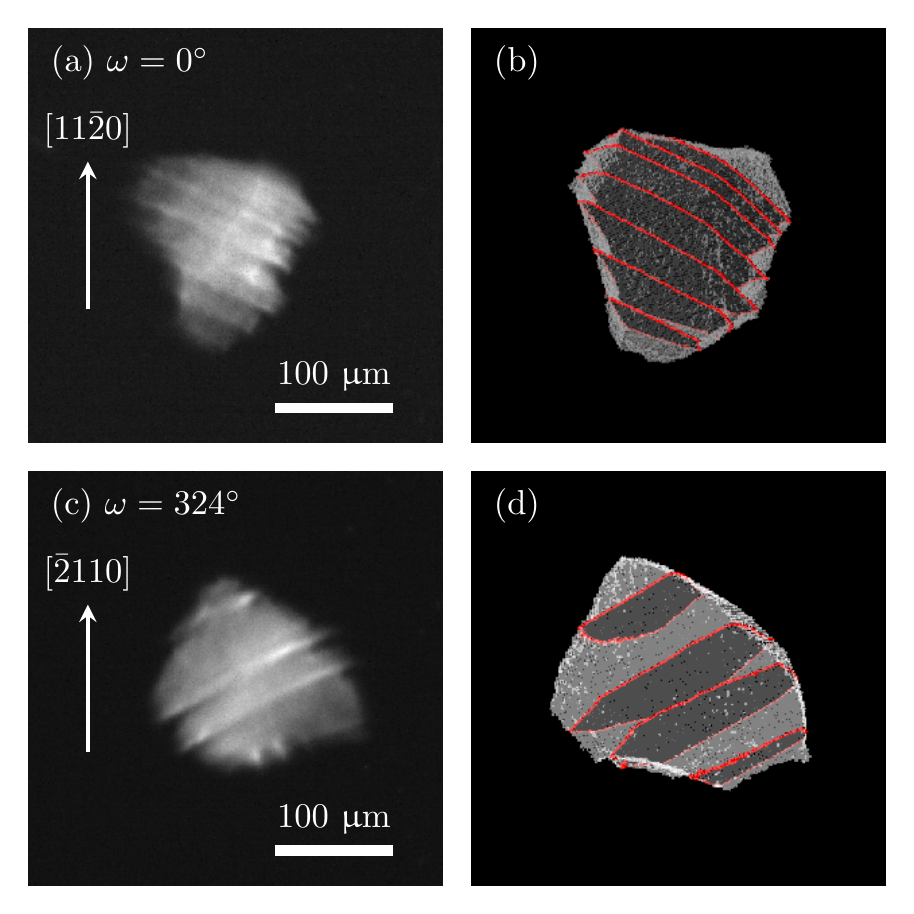}};
\end{tikzpicture}
\caption{Examples of slip planes identified by comparison between the recorded topographs (a and c for grains 572 and 579 respectively)  and the forward modeling of the diffraction for the corresponding grains (b and d), the slip plane intersections with the grain boundary are marked in color.}
\label{fig:ts3_tt_ex}
\end{figure}

The 75 topotomography scans were each post-processed to detect slip activity. The grain shape, position, and the 4 goniometer rotation angles were used to produce a simple forward simulation (parallel projection of the 3D grain volume) in the topotomography configuration. Using the grain orientation, slip planes are added in the volume of the grain by manually selecting the best match with the orientation contrast visible in the topographs over the \SI{360}{\degree}. If necessary, the active slip system was then identified using the highest Schmid factor within the plane. This is a tedious but efficient way of identifying slip planes that have produced plasticity in all the grains \cite{Proudhon2018}. \autoref{fig:ts3_tt_ex} depicts two topographs close to edge-on configuration (i.e. diffracted beam co-planar to slip plane) with their slip system identified and instantiated in the simulated projection images on the right figure panels. This process was repeated for the whole set of 75 topotomography scans comprising 55 grains.

\subsection{Orientation spread plots from topotomography}
\label{sec:ts3_tt_orientation_spread}
\begin{figure}
    \centering
    \includegraphics[width=1\textwidth]{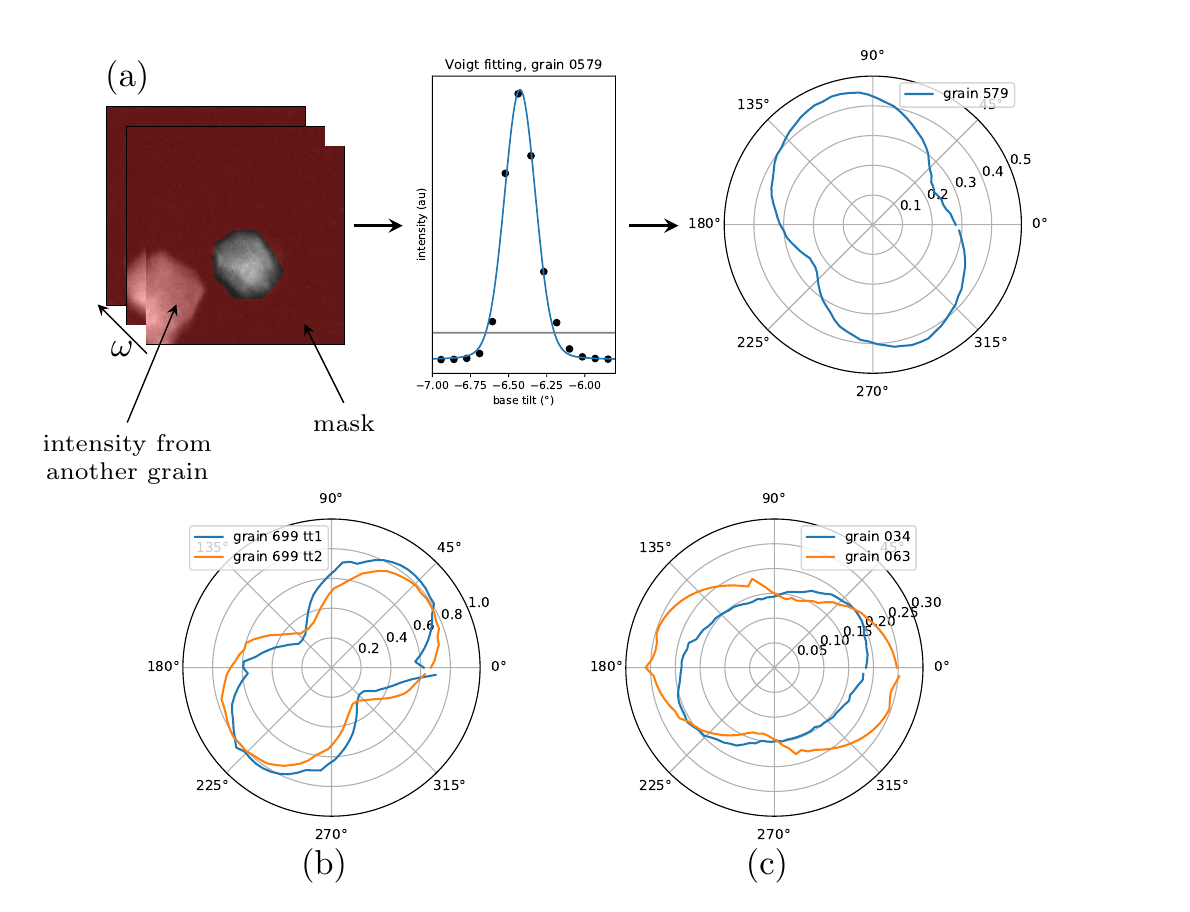}
    \caption{Orientation spread plots. (a) for each $\omega$ angle the individual topographs are masked and the resulting integrated intensity is fitted to extract the width at 10\% of the peak; the evolution of this value with $\omega$ depicts a typical dumbbell shaped curve here shown for grain 579 located in the bulk. (b) comparison of the effective misorientation measured on the same grain with two different reflections. (c) effective misorientation plots for surface grains 34 and 63. Note that small spikes can sometimes appear on the curves when spurious diffracting grains are located within the actual grain mask.}
    \label{fig:OSplots}
\end{figure}

Whereas the projection topographs shown in \autoref{fig:ts3_tt_ex} are integrated over angular range covered by the base tilt $\Theta$, the individual images taken during this motion (termed a rocking scan) constitute a rocking curve for each particular angle $\omega$. The change observed in the Bragg condition here is largely dominated by the lattice rotation (compared to elastic lattice distortions) and can be directly attributed to the orientation spread around the base tilt axis. To quantify this, the full width of the rocking curve (simply integrating the intensity collected by the detector at this angle) at 10\% of the peak is measured for each position of $\omega$ (every \SI{4}{\degree}) \cite{Proudhon2018}. To compute this value reliably, a mask for the grain being imaged needs to be created, because grains with similar Bragg conditions can diffract and be captured on the detector during the rotation. This phenomenon happens frequently enough to prevent a simple sum of the intensity on the detector, even with a proper ROI applied. To work around this problem, a forward simulation routine of the diffraction imaging was developed with Pymicro \cite{Pymicro} to create a mask of the grain projected onto the detector using the DCT reconstruction and the value of all motor positions (see \autoref{fig:OSplots}a). The mask was dilated by 10 pixels to allow for small grain deformations (and hence image distortions), which are not accounted for in the 3D DCT grain reconstruction used as input to create the mask.

A representative set of orientation spread curves are shown in \autoref{fig:OSplots}. They almost all depict a characteristic dumbbell shape as already observed in a previous topotomography experiment \cite{Proudhon2018}. This shape is the signature of a crystal bent perpendicularly to the elongation direction (in $\omega$) and can be related to dislocation activity and the presence of geometrically necessary dislocations (GNDs) introducing lattice rotations in the grains \cite{Gueninchault2016}. This is consistent with a single slip dominated deformation state as observed in most of the grains. An in depth look at the data processing shows that most of the $I_\omega(\Theta)$ rocking curves display regular single peak shapes which are the signature of rather homogeneous rotation field while some other display double peak shapes characteristic of more heterogeneous distribution of the lattice rotation or even formation of a subgrain (although this specimen was scanned ex situ so the initial state in the bulk is unknown). In this experiment some of the grains have been measured in topotomography for the first time using several reflections. For instance \autoref{fig:OSplots}b shows the orientation spread plots for grain 699 obtained with reflections \hkl(1 0 -1 1) and \hkl(1 1 -2 2) and one can see that the results are in very good agreement. Using several reflections will be useful in the future when extracting the orientation field from the measurements \cite{Vigano_COSS_2020} (not done in this paper).

\section{Results}

\begin{figure}
    \centering
    \includegraphics[width=1\textwidth]{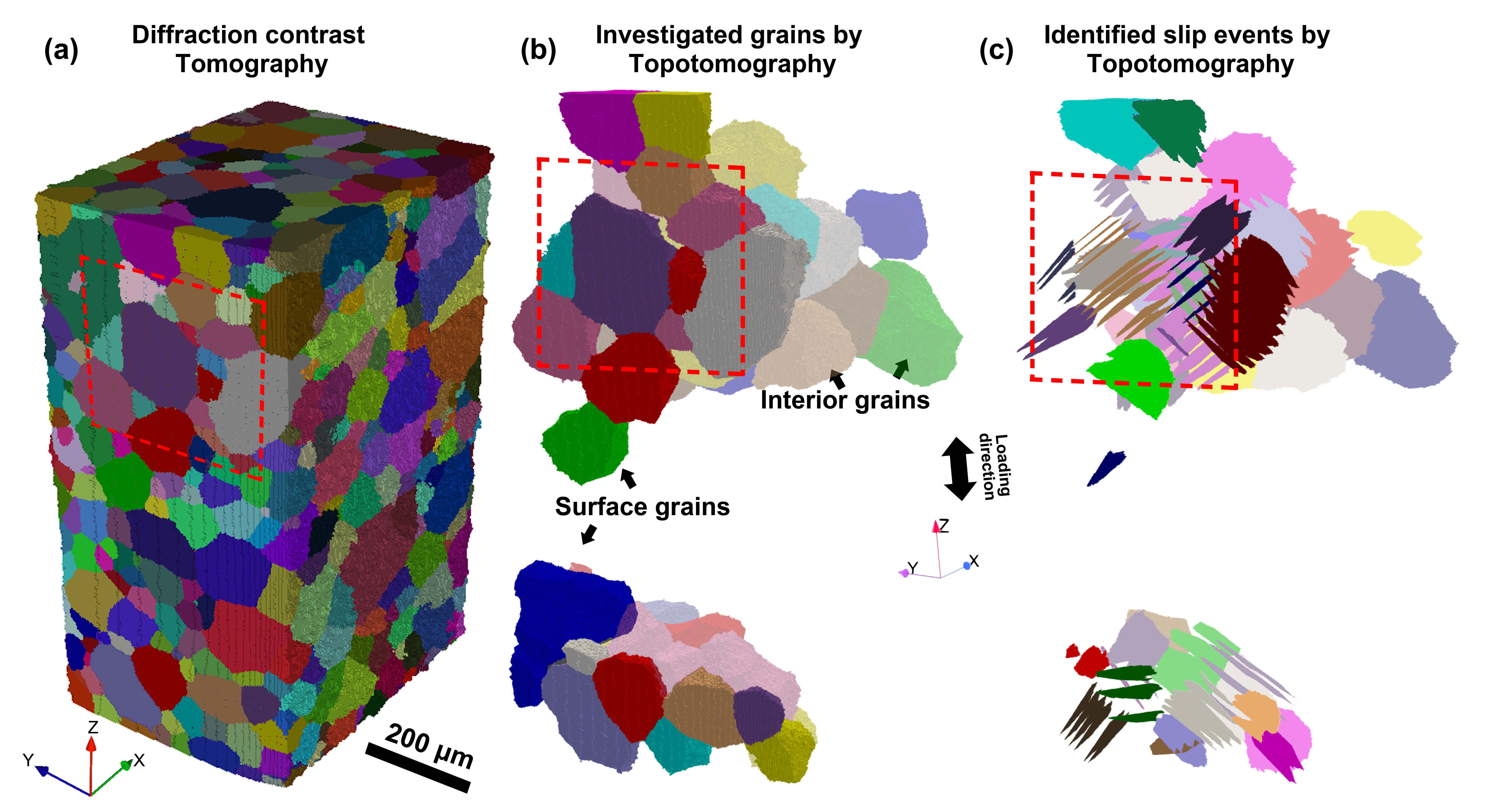}
    \caption{(a) DCT of the Ti-7Al specimen (random colors are used to differentiate grains), the front face corresponds to the free surface where the HR-DIC measurement was performed. (b) DCT of the grain investigated by topotomography. (c) Identification of surface and bulk slip events by topotomography. The red dashed box indicates the region highlighted in \autoref{DIC_ts3}(d-e).}
    \label{Topograins}
\end{figure}

The combination of X-ray diffraction contrast tomography and topotomography allow for the spatially resolved imaging of slip events within grains and with regard to the full 3D grain structure, as shown in \autoref{Topograins}. Reconstruction of the entire specimen gauge section by DCT is shown in \autoref{Topograins}(a). When compared to the EBSD map, the DCT reconstruction appears to be very accurate, including the shape of the surface grains. A detailed analysis showed that all of the 143 surface grains are captured in the EBSD map, except 6 very small and presumable shallow grains. On the other hand, the grains located close to the edges of the lateral faces that are difficult to capture in EBSD (slightly rounded specimen shape due to the polishing) are correctly reconstructed by DCT. The grains investigated by topotomography are displayed in \autoref{Topograins}(b). Two regions of surface grains were selected from HR-DIC measurements, with many grains exhibiting slip events after tensile deformation. The grains selected for topotomography measurements from within the bulk were chosen to be in contact with the selected surface grains and to have at least one low index lattice plane within reach by the diffractometer tilt motions (see Section \ref{sec:ts3_tt_setup}). Analysis of the topographs and the spatial correlation of DCT and topotomography measurements allows slip event detection and the determination of their location in the 3D grain structure, as shown in \autoref{Topograins}(c), for all the grains investigated by topotomography.

\subsection{Comparison between HR-DIC and Topotomography measurements}

Most of the grains investigated by topotomography (both surface and bulk grains) show slip events as observed by HR-DIC measurement. In addition, most of them appear to be in a single slip condition (single slip system activated).

\begin{figure}
    \centering
    \includegraphics[width=\textwidth]{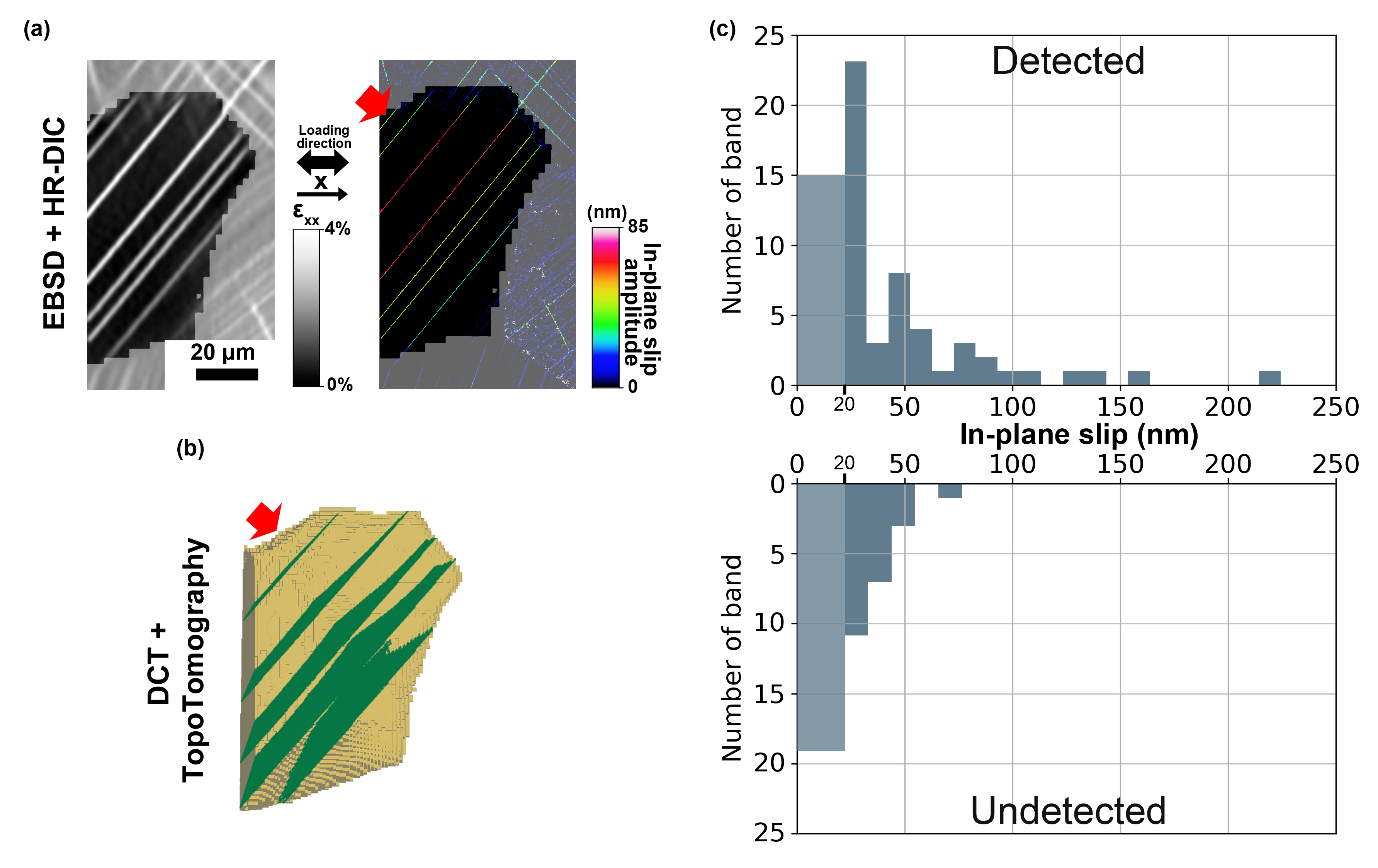}
    \caption{Comparison between HR-DIC and topotomography measurements: (a) strain map after deformation of a Ti-7Al specimen at about 0.8\%; associated in-plane slip amplitude map for a grain of interest. The in-plane slip amplitude corresponds to the physical in-plane displacement produced by a slip event onto the specimen free surface. (b) Recorded topographs and reconstructed grain shape and slip events for the grain of interest. (c) Distribution of slip events that are detected (top) and undetected (bottom) by topotomography measurements. The strain maps are used as reference to determine the location and number of slip events. The red arrows indicate two slip events at the edge of the grain of interest that are not detected by topotomography measurements. }
    \label{fig:DIC_detection}
\end{figure}

The HR-DIC strain measurements from the surface grains were analyzed completely independently from the topotomography measurements of the same grains, which contain full 3D grain information, in order to minimize potential bias and to validate the datasets with each other. Among the 55 investigated grains, 19 are surface grains. The HR-DIC and X-ray measurements are compared for all of these surface grains - containing 109 HR-DIC measured slip events. Specifically, the location and average in-plane slip amplitude of individual slip events (HR-DIC measurements) were extracted. An example is given in \autoref{fig:DIC_detection}(a and b) for a single grain with 7 intense surface slip events measured by HR-DIC, while only 5 slip events are detected by topotomography. The slip locations detected by topotomography in all the surface grains correlate accurately with slip events detected by HR-DIC measurement. For all the investigated surface grains, 67 slip events were detected by both techniques, while 41 were not detected by topotomography. The distributions of these detected and undetected events by topotomography are reported in \autoref{fig:DIC_detection}(c) as a function of their average in-plane slip amplitude, as obtained from the HR-DIC measurements. It must be noted that slip events with an in-plane slip amplitude below \SI{20}{\nano\meter} can not be differentiated using HR-DIC due to the resolution limit of measuring in-plane slip amplitude with this technique. The slip events that were undetected by topotomography, but detected by HR-DIC, are mainly low intensity slip events of in-plane slip amplitude below \SI{50}{\nano\meter} (or roughly lower than 170 emitted dislocations \cite{STINVILLE2020172}). Further inspection indicates that the most intense undetected slip events by topotomography are located near the edges of the grains as displayed at the red arrows in the grain in \autoref{fig:DIC_detection}(a and b). The detection of slip events at grain boundary is not favored using topotomography for two reasons: first, the diffracting grain volume at the periphery of a grain may be small and tend to have a lower diffraction signal to noise ratio. Second, the edges of the grains are more likely to accumulate high lattice rotations during deformation, which could obscure the contrast resulting from well defined slip plane locations and prevents their detection on the topographs.

\subsection{Contrast at slip events from topotomography measurement} \label{sec:contrast}

\begin{figure}
    \centering
    \includegraphics[width=\textwidth]{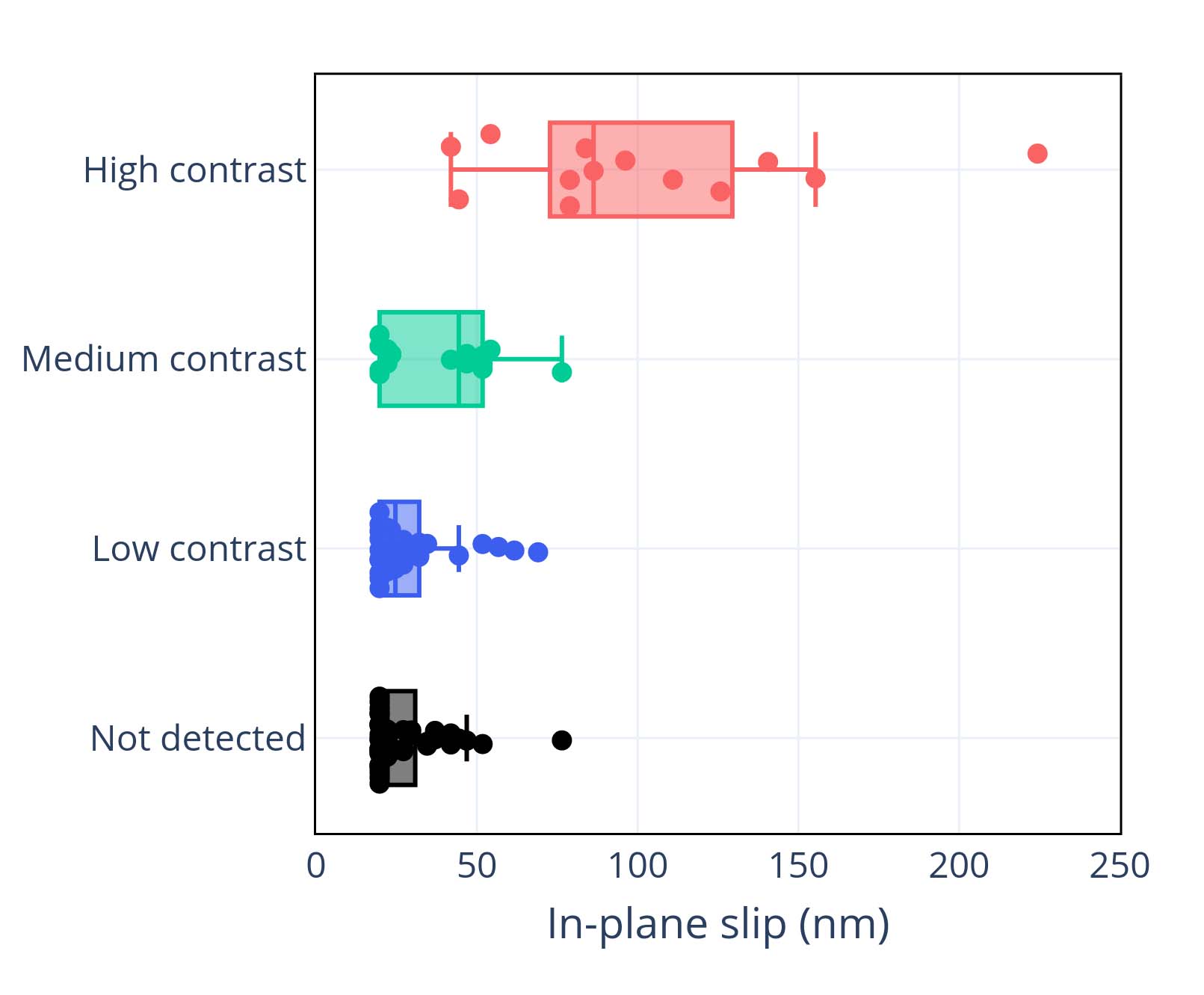}
    \caption{Contrast level of slip bands measurements by topotomography as a function of the HR-DIC surface measured in-plane slip intensity of the identical bands. Only slip events at the surface, or within the grains at the surface, are considered in the present analysis. The HR-DIC strain maps are used as a reference to determine the location and number of slip events. The slip bands that are not detected by topotomography measurements, but are present in the HR-DIC strain maps are labeled as "Not detected".  }
    \label{fig:TT_detection}
\end{figure}

While HR-DIC measurements provide the location and quantitative measurement of the slip events (in-plane slip amplitude given in \SI{}{\nano\meter}), topotomography measurements provide the slip location and an associated contrast on the topographs with contributions from the entire volume of the grain. The contrast detected in the topographs associated with slip events in the HR-DIC measured surface grains were classified into 4 groups: no contrast (no slip event detected by topotomography measurement but a slip event detected by HR-DIC), low contrast (slip event detected by topotomography with weak contrast on the topographs), medium contrast (slip events detected by topotomography with average contrast on the topographs), high contrast (slip events detected by topotomography with intense contrast on the topographs). The distributions of the in-plane amplitude as measured from HR-DIC measurements of slip events for these 4 groups are displayed in \autoref{fig:TT_detection}. 

As a general trend, the visibility of the topographic image contrast associated with a slip event is related to the in-plane slip amplitude of the slip event. High in-plane slip amplitude slip events tend to have high visibility in the topographs. Conversely, low intensity slip events tend to display low contrast in the topographs.

\subsection{Comparison between EBSD and Topotomography measurements}

Measurement of the lattice rotation magnitude induced during deformation has been performed for surface grains by conventional EBSD measurements and topotomography using grain orientation spread (GOS). The spread is defined as the average deviation between the orientation of each point in the grain and the average orientation of the grain. The conventional EBSD measurements provide information from the surface and near-surface of the grains, while topotomography measurements contain a measure of GOS (projections of the grain orientation distribution along the selected scattering vector) for the entire volume of the grain. Note that lattice rotations around the scattering vector are not captured by the rocking curve analysis presented in the current work.

\begin{figure}
    \centering
    \includegraphics[width=\textwidth]{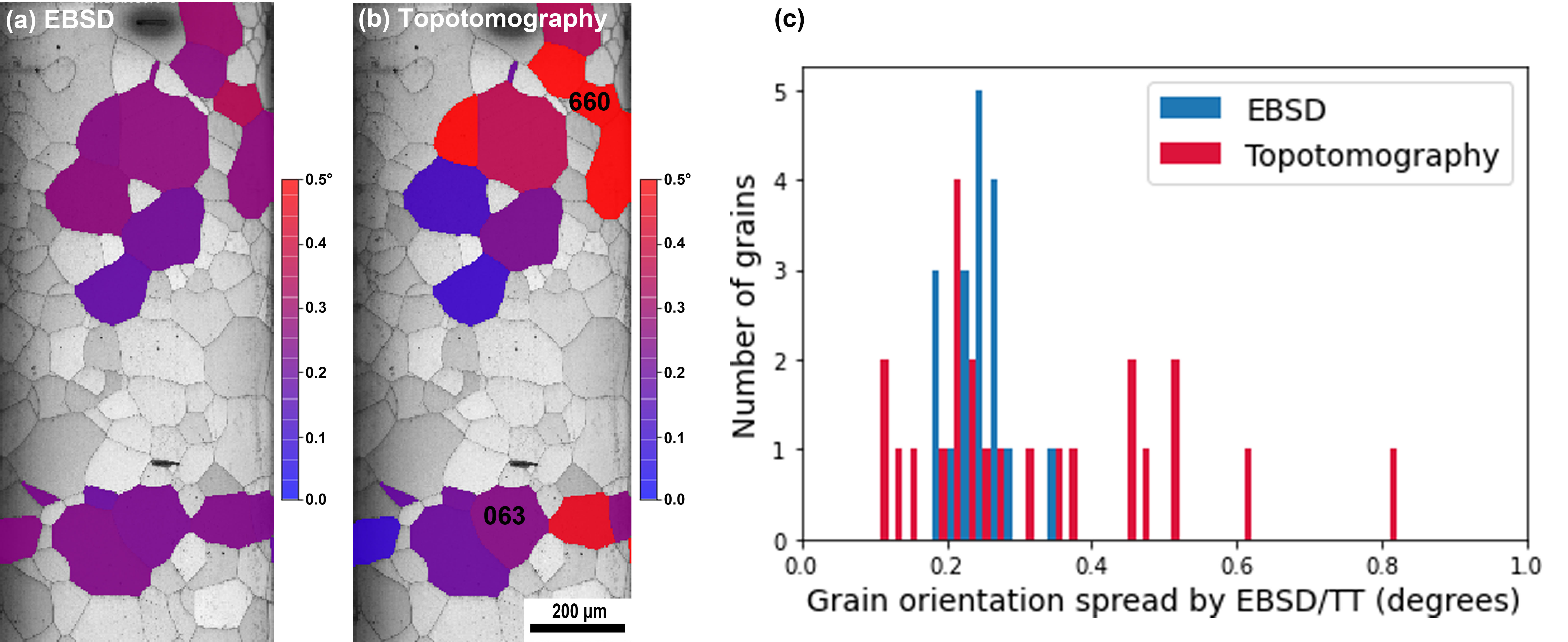}
    \caption{Grain Orientation Spread (GOS) for the investigated surface grains from conventional 2D EBSD measurements (a) and full 3D topotomography (b) measurements. The distribution of the (GOS) is displayed in (c). }
    \label{fig:spread}
\end{figure}

A GOS map is shown in \autoref{fig:spread}(a) from EBSD alongside measurements of the orientation spread made by topotomography. The map in \autoref{fig:spread}(b) was constructed by overlaying the EBSD image quality signal (to provide a sensible background) onto a map of the orientation spread measured by topotomography (see Section \ref{sec:ts3_tt_orientation_spread}, and captures the orientation spread from the entire 3D grain structure. The distribution of the orientation spread obtained from both techniques is displayed in \autoref{fig:spread}(c).

%Note that the two quantities are not identical but almost equivalent in nature. In EBSD, the GOS is computed as the mean of the misorientation of all pixel of a given grain with respect to the mean grain orientation.

%One main difference is that with EBSD, only the extreme surface is probed, whereas with topotomography, the whole grain contributes to the measurement.

The analysis from EBSD and topotomography of the orientation spread in surface grains shows overall similar values, indicating that the measurement of orientation spread by topotomography is sensitive enough to capture the evolution of lattice rotations during plastic deformation. However, the grain-to-grain GOS variation in EBSD (surface measurement) compared to topotomography (entire 3D grain) is different, indicating the significance of the bulk measurement information.  The observed differences can be attributed partly to the lower angular resolution of EBSD and partly to the orientation of the active slip direction with respect to the sample surface, this is discussed in more detail in Section \ref{sec:lattice_rot_discussion}.

\section{Discussion}

Plastic deformation does not occur uniformly during the loading of polycrystalline metallic materials, but instead it occurs initially by localized slip in regions where dislocations first overcome obstacles to deformation. This manifests at the surface of the specimen by the formation of surface steps \cite{Mitchell1993}. With continued straining, numerous occurrences of these slip events result in a continuous network of plastic flow across the entire cross section of the specimen, corresponding to macroscopic yielding of the specimen \cite{Echlin2021}.

Plasticity and its transmission across grains has been extensively investigated at the surface of the specimen during loading. However, very little experimental data is available on the localization and transmission of plasticity in the material bulk as a function of the 3D grain structure. For example, it has recently been observed that the 3D grain structure is especially critical to understanding the origin of incipient localization in nickel-based superalloys \cite{CHARPAGNE2021117037}. The combination of DCT and topotomography \cite{Ludwig2007,Proudhon2018} provides the unique capability to detect slip events as a function of the grain structure. This technique has tremendous opportunity for investigation of the bulk slip localization, however the sensitivity of the technique with regard to slip detection has not yet been quantified and parameters other than the slip amplitude may influence the visibility of slip localization in topographs.

\subsection{X-ray Topotomography slip event detection and slip event contrast}

\begin{figure}
    \centering
    \includegraphics[width=\textwidth]{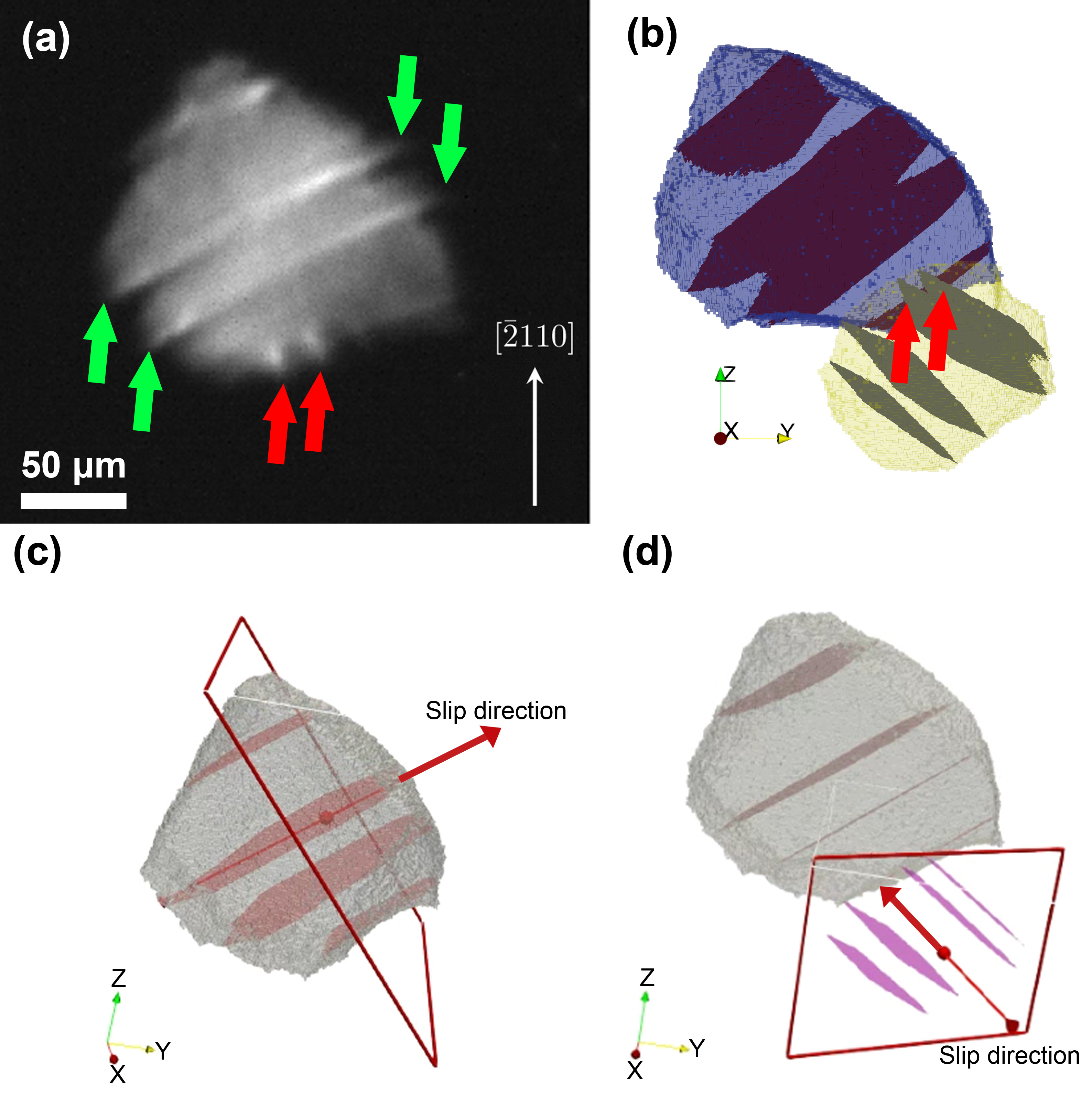}
    \caption{(a) Topograph of an investigated grain with intense contrast at slip events (green arrows) and impingement of slip events at the boundary from a neighboring grain (red arrows). (b) 3D DCT reconstruction of the investigated grain (blue) and a neighboring grain (yellow) and the overlaid slip event planes, as detected by topotomography. (c-d) Slip directions of the active basal slip systems for the investigated grain and neighboring grain. }
    \label{fig:grain659_579}
\end{figure}

Compared to a previous study in an AlLi alloy \cite{Proudhon2018}, the level of deformation is higher in the presently investigated Ti-7Al sample (0.8\% vs 0.3\%), as seen in the integrated topographs in \autoref{fig:ts3_tt_ex} and \autoref{fig:grain659_579}(a). Due to the higher level of deformation, the grain projections formed on the topographs are more deformed - making it more challenging to identify the slip planes. Nevertheless, the vast majority of the grains depict strong contrast localization in the form of bands, just as in the AlLi experiment \cite{Proudhon2018}. Using the grain orientations from DCT, these bands can be correlated with active slip planes in the bulk of the grain. Another difference in the Ti-7Al material is that the orientation contrast localized around slip bands remains visible on a much larger angular range than in the previous AlLi study (typically \SI{50}{\degree} vs \SI{10}{\degree}). In summary, using this slip system identification technique, it appears that we approach the upper limit (i.e. 1 or 2\% plastic strain) for the topotomography technique. At increased levels of plasticity, the projection contrast becomes too convoluted to reliably identify the slip planes. This limitation restricts the use of topotomography for investigation of the slip activity at relatively low levels of plastic deformation. However, we note that for fatigue properties - typically much of the activity is dictated by the first few cycles \cite{STINVILLE2020172} at relatively low strain - potentially making topotomography an extremely useful technique. 

%The effects of these chemical and microstructural attributes on the localization of plasticity by slip is generally investigated by surface measurements, such as optical, atomic force or scanning electron microscopy based measurements, which provide limited insight to the complex 3D microstructure that exists below the surface. . X-ray synchrotron based measurement techniques are of great interest for the detection of bulk plasticity in metals and the consequential effect of the 3D microstructure (particularly grain structure) on the mechanical properties.

%The combination of X-ray tomography and topotomography \cite{Ludwig2007,Proudhon2018} provide the unique capability to detect slip events as a function of the grain structure. This technique as tremendous opportunity for investigation of the bulk slip localization, however the sensitivity of the technique with regard to slip detection is not well defined. 

The results in Section \ref{sec:contrast} demonstrates that slip events with low intensity as detected by HR-DIC measurements were not all detected by topotomography, indicating that low intensity slip events do not produce enough contrast to be detected integrated topotomography images. Note that in this paper we have used integrated images to detect slip events while one could also use individual images (typically 64 per topograph here) to look at contrast variations (hardly tractable with a manual approach as pursued here but with potentially increased sensitivity if automated). Topographic image contrast can be roughly divided into two categories: (i) contrasts related to dynamical diffraction and extinction effects and (ii) orientation contrast linked to local variations of the effective misorientation (i.e. the combined effect of rotation and elastic distortions of the crystal lattice) \cite{Tanner1976}. Whereas the former can be predominant in nearly perfect single crystals, the latter are more relevant in the case of metal grains, typically containing a high density of dislocations, even in the as-recrystallized state of the material. Topographic image contrast of slip localization events can be attributed to local gradients in lattice rotation induced by high dislocation densities generated by the avalanche and pile-up processes associated with slip events. These processes are well investigated in the literature \cite{Alcala2020,GUO2015229,Villechaise_2012}. The interaction between dislocation pile-ups and grain boundaries gives rise to heterogeneous stress distributions. Such stress heterogeneity leads to very local gradients in effective misorientation of the crystal lattice \cite{GUO2015229,Villechaise_2012,Polcarova2006} that are captured by topotomography measurements. Other effects, like local reduction of the scattered intensity due to the disorder in the crystal lattice in proximity of slip localizations are likely to contribute to the intensity variations observed in the topographs.

To further investigate the characteristics of the contrast at slip events in the topograph, a specific grain in the specimen bulk with significant topograph contrast intensity is displayed in \autoref{fig:grain659_579}. The topograph from the tilt series with the highest contrast intensity is presented in \autoref{fig:grain659_579}(a). Four slip events within the investigated grain are observed and depicted by the purple planes in the reconstruction of the grain (blue colored) and slip events in \autoref{fig:grain659_579}(b). The contrast intensity varies along the plane of the slip events and is especially high on the left and right edges, as indicated by green arrows. Interestingly, intense contrast is also observed at the bottom edge of the grain (at the red arrows in \autoref{fig:grain659_579}(a)), which correspond to the impingement of two slip bands that developed in the neighboring grain (yellow grain in \autoref{fig:grain659_579}(b)). 

Both slip systems in the investigated grain (blue) and neighboring grain (yellow) were identified as basal type. The slip direction for the active basal slip systems in the detected slip planes for both grains are depicted with red arrows in \autoref{fig:grain659_579}(c and d). The high contrast intensity occurs near the grain edges (green arrows in \autoref{fig:grain659_579}(a)) in the topograph where the slip direction is orthogonal to the imaging plane. In other words, the intensity of the contrast is maximized when the slip direction is in the plane of the page, as approximated by the red arrow in \autoref{fig:grain659_579}(c). Furthermore, the two intense points of contrast at the red arrows in \autoref{fig:grain659_579}(a), which are due to impingement of the intense slip bands from the neighboring grain (yellow grain in \autoref{fig:grain659_579}(b)), occur at the location where the active basal slip system direction is pointing towards the grain boundary. Identical types of observations were made at several instances of grains and slip events in the topotomography Ti-7Al dataset. 

From the previous example, it is observed that some of the high contrast regions in topographs that reveal the presence of slip events, originate from the pile-ups of dislocation at grain boundaries. These pile-ups either induce significant contrast in the grain where the slip events develops or in the neighboring grain where slip band impingement occurs. The lattice rotation contrast induced by dislocation pile-ups have previously been investigated by HR-EBSD measurements in titanium alloys \cite{GUO2015229}, and their amplitudes are on the order of the magnitude that corresponds to the contrast observed on topotomograph. It is therefore reasonable now to consider than the contrast observed in topographs that provide the identification of slip events may be enhanced in proximity of grain boundaries (due to pile-ups). On the other hand, the contrast also appear to come from the collective presence of dislocations (alongside  their stress fields) moving within the slip band and providing edge-on contrast. This consideration can explain the propensity for the topotomography technique to tend to not detect low intensity slip events, which are associated with low dislocation densities at pile-ups and where the contrast created by the band within the grain is too faint. An in situ experiment would be helpful to analyse the evolution of contrast when a slip band and its pile-ups increase in intensity as the deformation progresses.

\subsection{Slip transmission in the bulk}

\begin{figure}
    \centering
    \includegraphics[width=\textwidth]{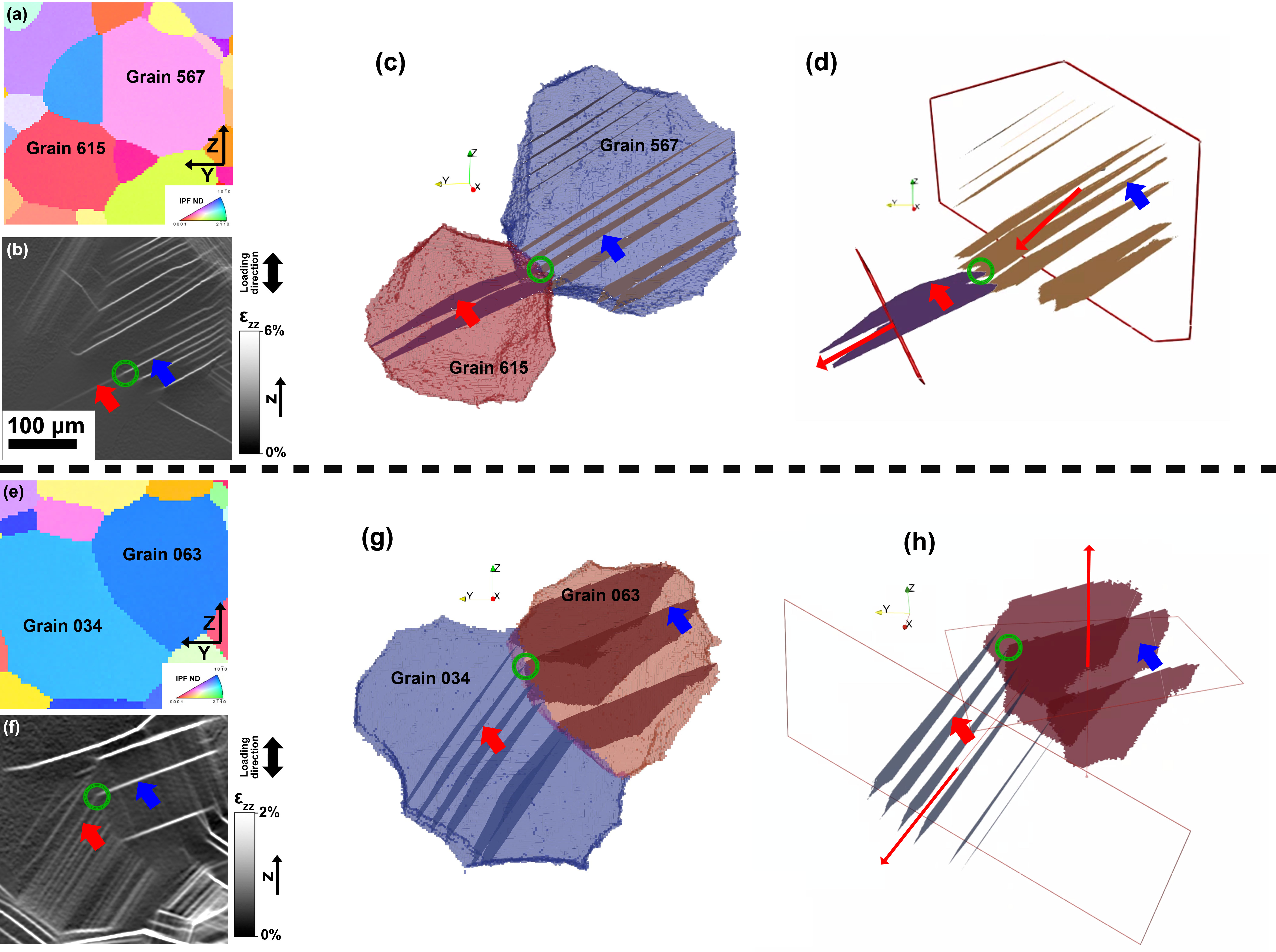}
    \caption{Two transmission events observed by DCT and topotomography. (a,e) EBSD map of the surrounding grain structure around the transmission events. (b,f) HR-DIC $\epsilon_{zz}$ strain map showing the investigated slip transmission events. (c,g) 3D reconstruction of the grain structure and slip events for the grains of interest. The incoming and transmitted slip are indicated by blue and red arrows, respectively. The green circles highlight the location of slip transmission. (d,h) 3D reconstruction of the active slip planes detected by topotomography for the grains of interest. }
    \label{fig:Transmission_1}
\end{figure}

One of the interesting applications of topotomography is the investigation of slip transmission. Typically, transmission analyses are performed from surface measurements that fail to experimentally capture the inclination of the slip events and geometry of the grain boundaries in the sub-surface/bulk. While slip transmission in the bulk of the specimen will be more rigorously studied in a future study of Ti-7Al, demonstrative examples are presented in \autoref{fig:Transmission_1} from regions measured by HR-DIC and topotomography, where slip events (blue arrows) are transmitting in the neighboring grains (red arrows). The incoming and transmitted slip events were detected in the topotomography measurements and their 3D representation is provided in \autoref{fig:Transmission_1}(c and d) for the first example and in \autoref{fig:Transmission_1}(g and h) for the second example. Incoming slip events (blue arrows) in both of the investigated grains were identified as the slip system with the highest Schmid factor, and the active slip directions are depicted by the long red arrows in \autoref{fig:Transmission_1}(d and h). The associated HR-DIC and EBSD maps are provided for both examples in \autoref{fig:Transmission_1}(a,b and e,f). 

The first example shown in \autoref{fig:Transmission_1}(a-d) displays a direct transmission event. Upon consideration of the Schmid factor for grain 615, slip system %\hkl(-1 0 0)[0 1 0]
\hkl(-1 0 1 0)[-1 2 -1 0] has the highest Schmid factor of 0.405 and is not activated, while the slip system %\hkl(-1 1 0)[1 1 0] 
\hkl(-1 1 0 0)[1 1 -2 0] is activated (purples plane in \autoref{fig:Transmission_1}(c-d)) with a lower Schmid factor of 0.373. Consequently, the transmission event triggered a slip system in grain 615 with a lower Schmid factor but with favored transmission configuration. The $m'$ factor \cite{Luster1995} is usually an effective indicator to describe the propensity of a incoming slip event to transmit to the neighboring grain. It is defined as $m'= (d_1\cdot{}d_2)(n_1\cdot{}n_2)$, with $d_1$ the slip direction of the incoming slip system, $d_2$ the slip direction of the outgoing slip system, $n_1$ the slip plane normal of the incoming slip system and $n_2$ the slip plane normal of the outgoing slip system. $m'$ is ranging between 0 and 1 and high values indicate that better geometric compatibility exists between both slip systems that favor slip transmission. The $m'$ factor between the incoming and transmitted slip event is 0.7 while the one between the incoming slip and the highest Schmid factor slip system in grain 615 is 0.1. This is evidence of plasticity behavior that is dependent on the local surrounding microstructural configuration as has previously been reported for titanium alloys \cite{HEMERY2018277,BIELER2014212}. The topotomography measurements provide accurate spatial resolution to capture such behavior at the surface as conventional surface measurements (trace analysis from SEM, HR-DIC or AFM), but also into the specimen bulk.  

\begin{figure}
    \centering
    \includegraphics[width=0.65\textwidth]{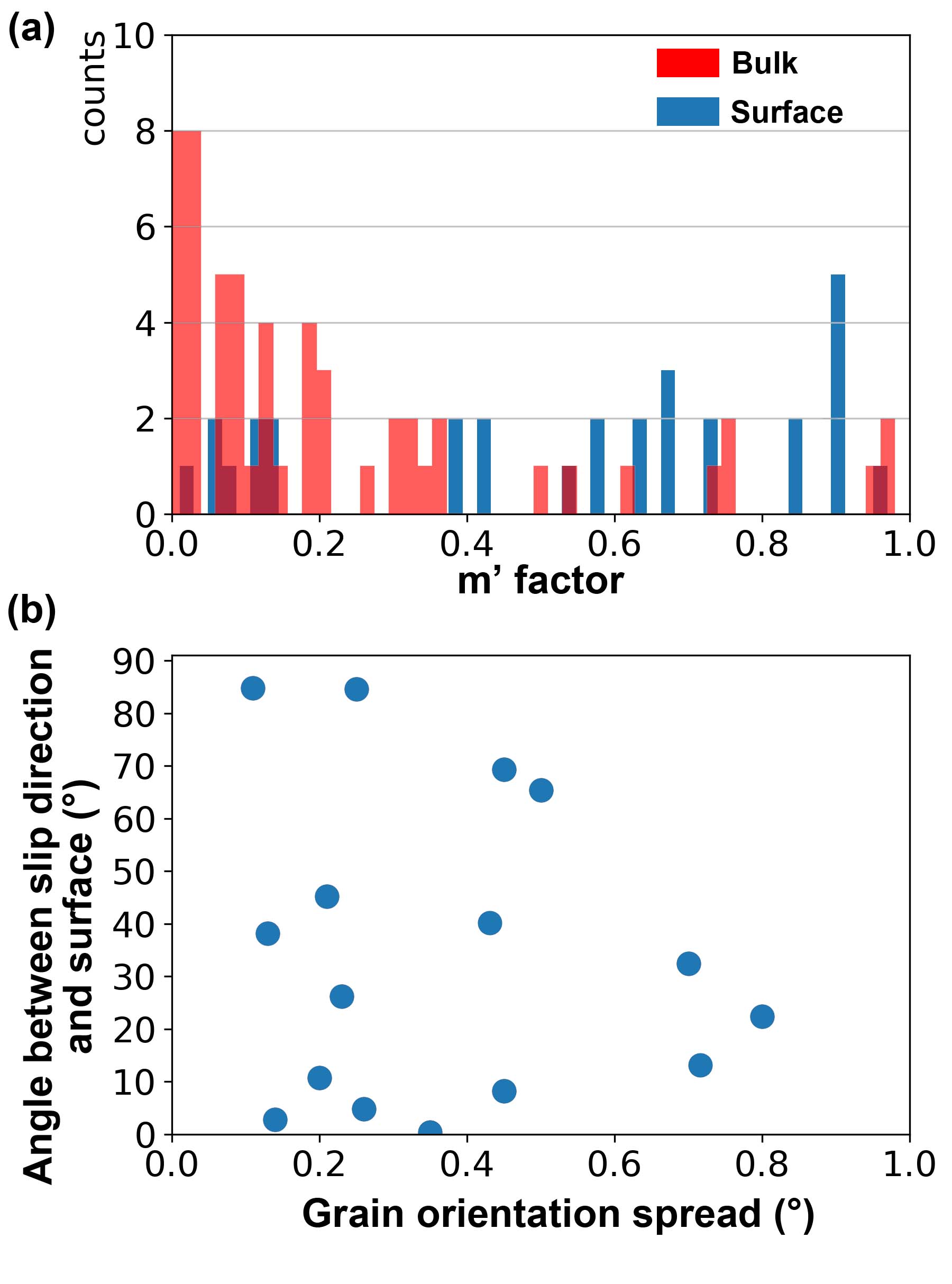}
   \caption{(a) Distribution of m' factor between surface grains and from grains in the bulk. The m' factors were calculated for all slip events detected by topotomography that are connected (at least one voxel in common at a grain boundary) across surface grains and from bulk grains. (b) Relation between GOS from topotomography measurements and the slip direction in relation to the free surface.}
    \label{fig:mtransmission}
\end{figure}

Interestingly, the topotomography measurements allow differentiation between slip activity that occurs at the surface and in the bulk. The m' factors were calculated for all slip events that are connected (at least one voxel in common at a grain boundary) across surface grains and from bulk grains. The distributions of the $m'$ factor is reported in \autoref{fig:mtransmission}. It is striking to notice such a difference in the distribution from bulk and surface grains. Surface grains tend to promote slip transmission and as a consequence, the $m'$ factor is important when investigating slip activity at the surface, as reported in the literature \cite{HEMERY2018277}. Conversely, the distribution of the $m'$ factor between bulk grains indicates that it may not be as relevant for the description of slip activity in the bulk. Direct transmission may not be of significant importance within the bulk, in agreement with the recent observation of the importance of the triple junction within the bulk \cite{CHARPAGNE2021117037}. In this work, it was observed in a deformed metallic material that a large amount of the slip events that develop originate from triple junction lines that have significant stresses. From the present analysis, the possibility should be considered that slip activity within the bulk is mainly controlled by stress heterogeneity at junction lines, with slip transmission having a lesser influence.

Interestingly, multiple peculiar cases of direct transmission were observed at the surface of the specimen as displayed by the second example in \autoref{fig:Transmission_1}. In this example, the incoming slip event and the transmitted slip event are connected by the unique surface point highlighted by green open dots in \autoref{fig:Transmission_1}(c and d). The transmission factor m' is calculated as 0.1 indicating a configuration that does not favor slip transmission since both the plane normal and slip directions of the active slip systems are highly disoriented from each other, as displayed in \autoref{fig:Transmission_1}(d). From the 3D observation of this slip transmission event, and considering that the slip direction of the incoming slip event is almost normal to the surface, it can be hypothesized that the intense incoming slip event developed a relatively significant surface step that triggered slip activation in the neighboring grain. The constraint of the slip extrusion at a grain boundary can generate large local stresses that then can trigger slip in the neighboring grain. This kind of transmission event was observed in several instances in the investigated surface grains and contributed to the low values of the m' factor in the distribution in \autoref{fig:mtransmission}(a). It is generally accepted that direct or indirect slip transmission is controlled by dislocation transmission across grain boundaries \cite{Luster1995}. However, the present example evidences another type of transmission related more likely to stress concentration and surface effects due to the constraint of a slip step near grain boundary at the free surface. This is further evidence that the m' factor is not always relevant for characterization of slip transmission at the surface.

\subsection{lattice rotation from bulk measurements}
\label{sec:lattice_rot_discussion}
Significant differences in orientation spread, induced by deformation, are observed from near-surface measurements (EBSD) and bulk measurements (topotomography). This result is expected since the interior of the grain can evolve differently during deformation than the near surface of the grain. It is important to question the relationship between slip and orientation spread from near-surface and bulk measurement. 

For instance, the grain labeled 660 has an orientation spread of \SI{0.5}{\degree}, whereas grain 063 displays a spread of only \SI{0.25}{\degree}. The analysis of the active slip system Burgers vectors shows that in grain 063, the vector points out of the free surface meaning the dislocations can readily escape the volume. However, in grain 660 the vector is almost parallel to the surface meaning that the dislocations do not escape and pile up at the grain boundary, storing GNDs and creating mosaicity as a result. This behavior appears to be general for the surface grains measured by topotomography. This relation is evidenced in \autoref{fig:mtransmission}(b) where the orientation spread is displayed as a function of the angle between the slip direction and the surface vector of the active slip plane. Grains with high orientation spread tend to display a slip direction that is close to the plane of the free surface. This provides more evidence of the effect of the free surface on the plastic activity. Other factors such as the grain size, the level of plasticity, the neighboring grains and the grain shape are also certainly relevant and are required to be investigated to highlight the relation between orientation spread and plastic activity. 

The relationship observed previously between orientation spread and slip direction is not observed in orientation spread for the near-surface measurements obtained by EBSD. This indicates that EBSD and other surface measurements may not be representative of the plasticity that occurs within the entire grain, highlighting an advantage of the topotomography measurements, even for surface grains.

%One exception is grain 151 (slip direction points toward the surface so that we might expect a lower lattice rotation). As pointed out in section \label{sec:results} a small shallow grain visible in EBSD was not captured in the DCT measurements and it it plausible that this grain blocks the escape of the dislocation resulting in large mosaicity ($1.4^\circ$ in this grain).

\section{Future Directions}

Topotomography is a material characterization technique the has matured over the last 10 years, all while becoming more and more automated. It now also benefits from major synchrotron source upgrades that reduce data collection times to a few minutes per grain, as compared to several tens of minutes for the experiment reported here. This enables the systematic mapping of massive regions of grains, selected from an initial DCT reconstruction of the grain volume. Scripts performing automated TT grain alignment and driving scan sequences (DCT, PCT and series of TT scans over a list of grains) have been developed and enable repeated acquisitions at increasing levels of applied strain, which could be instrumental in capturing plasticity propagation events. This framework and in particular the selection of candidate grains could possibly be guided by a digital-twin mechanical analysis conducted on the initial DCT reconstruction.

The question of the nature of the contrast mechanism(s) remains open, with likely origins being orientation contrast created by GNDs piling up at grain boundaries as evidenced in this paper, dislocations structures located within the slip bands, or a combination of both. Dislocation dynamics simulations based on experimental microstructures and coupled with forward modeling diffraction is currently being developed and may accurately inform on the contrast mechanisms in the near future. 

Another current challenge relies in inverting the topotomography images to reconstruct the orientation field. Since TT can not resolve lattice rotations around the probed scattering vector, the inversion has to include complementary grain projection data from DCT. This is in principle feasible using iterative approaches as demonstrated in \cite{Vigano_COSS_2020} or by adapting the forward modeling strategy developed by \cite{Suter2006} to the combined DCT and TT acquisition geometry. As the topotomography geometry is more complex (it involves several rotations around the eucentric point), the sensitivity and spatial resolution of this approach may suffer from diffractometer error motion and requires an accurate calibration of the instrument geometry. Correlative experiments will be instrumental to validate the reconstructions in the future. To this end, recent updates in experimental stations at ESRF such as the nanobeam scanning 3DXRD station at ID11 or the dark field X-ray microscopy station at ID06 can be leveraged.

%\begin{table}[h!]
%\centering
%
%\begin{tabular}{|c|| c c c|} 
% \hline
% & Topograph & Grain (low def. - ?? rocking angle) & Grain (high def. - ? rocking angle) \\ [0.5ex] 
% \hline\hline
%Previous investigation \cite{Proudhon2018}  & min & min & min \\ 
% \hline
% Present investigation & min & min & min \\
% \hline
% New ESRF beam line & min & min & min \\
% \hline
%
%\end{tabular}
%\caption{Acquisition times improvement over the past iteration of the topotomography technique and syncrotron beam line.}
%\label{table:time}
%\end{table}

\section{Conclusions}

The correlative measurements from HR-DIC, topotomography and DCT made explicit the sensitivity and spatial resolution of the topotomography measurements. Topotomography measurements were able to detect most of the intense slip events that developed during deformation of a titanium alloy. Slip events of very low intensities or in regions of high lattice rotation were not detected by topotomography measurements, at least with the slip detection method used in this paper. Topotomography measurements provide the unique opportunity to investigate the plastic activity in a large number of grains both at the surface and in the bulk. Both slip events location and orientation spread are obtained from topotomography. Several transmission events were investigated in a titanium alloy by topotomography. The analysis of slip transfer between surface grains shows that many slip transmission events can be rationalized in terms of the geometrical alignment of activated slip systems in neighboring grains (m' factor for instance). However, the slip activity within the bulk of the specimen is observed to \textbf{not} be controlled by slip transmission in the same extent as those of surface grains. In addition, an unknown transmission phenomena was observed at the surface, indicating again that completely different plastic behavior may occur in the bulk versus at free surfaces.

\section{Acknowledgments}
The authors would like to thank ESRF for beam time allocation under proposal MA3921 and Chris Torbet for machining the Ti-7Al specimens. HP also like to thank Georges Cailletaud (holder of the Cristal chair funded by Mines ParisTech and Safran) for the financial support to visit UCSB. The authors also acknowledge the support of ONR Grant N00014-19-2129. The MRL Shared Experimental Facilities are supported by the MRSEC Program of the NSF under Award No. DMR 1720256; a member of the NSF-funded Materials Research Facilities Network (www.mrfn.org). PGC would like to acknowledge support provided by the Naval Research Laboratory under the auspices of the Office of Naval Research.

% References with bibTeX database:
\bibliographystyle{elsarticle-num}
\bibliography{bib.bib}

\begin{thebibliography}{10}
\expandafter\ifx\csname url\endcsname\relax
  \def\url#1{\texttt{#1}}\fi
\expandafter\ifx\csname urlprefix\endcsname\relax\def\urlprefix{URL }\fi
\expandafter\ifx\csname href\endcsname\relax
  \def\href#1#2{#2} \def\path#1{#1}\fi

\bibitem{Mitchell1993}
J.~Mitchell,
  \href{https://onlinelibrary.wiley.com/doi/abs/10.1002/pssa.2211350211}{Elementary
  processes in the formation of slip bands in single crystals of $\alpha$-phase
  cu-al alloys}, physica status solidi (a) 135~(2) (1993) 455--466.
\newblock \href
  {http://arxiv.org/abs/https://onlinelibrary.wiley.com/doi/pdf/10.1002/pssa.2211350211}
  {\path{arXiv:https://onlinelibrary.wiley.com/doi/pdf/10.1002/pssa.2211350211}},
  \href {http://dx.doi.org/https://doi.org/10.1002/pssa.2211350211}
  {\path{doi:https://doi.org/10.1002/pssa.2211350211}}.
\newline\urlprefix\url{https://onlinelibrary.wiley.com/doi/abs/10.1002/pssa.2211350211}

\bibitem{Lombros2016}
W.~Abuzaid, H.~Sehitoglu, J.~Lambros,
  \href{https://doi.org/10.1080/09603409.2016.1152421}{Localisation of plastic
  strain at the microstructurlal level in hastelloy x subjected to monotonic,
  fatigue, and creep loading: the role of grain boundaries and slip
  transmission}, Materials at High Temperatures 33~(4-5) (2016) 384--400.
\newblock \href
  {http://arxiv.org/abs/https://doi.org/10.1080/09603409.2016.1152421}
  {\path{arXiv:https://doi.org/10.1080/09603409.2016.1152421}}, \href
  {http://dx.doi.org/10.1080/09603409.2016.1152421}
  {\path{doi:10.1080/09603409.2016.1152421}}.
\newline\urlprefix\url{https://doi.org/10.1080/09603409.2016.1152421}

\bibitem{LIU2019260}
J.~Liu, N.~Vanderesse, J.~Stinville, T.~Pollock, P.~Bocher, D.~Texier,
  \href{https://www.sciencedirect.com/science/article/pii/S135964541930134X}{In-plane
  and out-of-plane deformation at the sub-grain scale in polycrystalline
  materials assessed by confocal microscopy}, Acta Materialia 169 (2019)
  260--274.
\newblock \href
  {http://dx.doi.org/https://doi.org/10.1016/j.actamat.2019.03.001}
  {\path{doi:https://doi.org/10.1016/j.actamat.2019.03.001}}.
\newline\urlprefix\url{https://www.sciencedirect.com/science/article/pii/S135964541930134X}

\bibitem{Stinville2015expmech}
J.~Stinville, M.~Echlin, D.~Texier, F.~Bridier, P.~Bocher, T.~Pollock,
  \href{https://doi.org/10.1007/s11340-015-0083-4}{Sub-grain scale digital
  image correlation by electron microscopy for polycrystalline materials during
  elastic and plastic deformation}, Experimental Mechanics 56~(2) (2015)
  197--216.
\newblock \href {http://dx.doi.org/10.1007/s11340-015-0083-4}
  {\path{doi:10.1007/s11340-015-0083-4}}.
\newline\urlprefix\url{https://doi.org/10.1007/s11340-015-0083-4}

\bibitem{BONNEVILLE200887}
J.~Bonneville, C.~Coupeau,
  \href{https://www.sciencedirect.com/science/article/pii/S0921509307008805}{Quantitative
  atomic force microscopy analysis of slip traces in ni3al yield stress
  anomaly}, Materials Science and Engineering: A 483-484 (2008) 87--90, 14th
  International Conference on the Strength of Materials.
\newblock \href {http://dx.doi.org/https://doi.org/10.1016/j.msea.2006.12.158}
  {\path{doi:https://doi.org/10.1016/j.msea.2006.12.158}}.
\newline\urlprefix\url{https://www.sciencedirect.com/science/article/pii/S0921509307008805}

\bibitem{AUBERT20169}
I.~Aubert, N.~Saintier, J.-M. Olive, F.~Plessier,
  \href{https://www.sciencedirect.com/science/article/pii/S1359645415300938}{A
  methodology to obtain data at the slip-band scale from atomic force
  microscopy observations and crystal plasticity simulations. application to
  hydrogen-induced slip localization on aisi 316l stainless steel}, Acta
  Materialia 104 (2016) 9--17.
\newblock \href
  {http://dx.doi.org/https://doi.org/10.1016/j.actamat.2015.11.042}
  {\path{doi:https://doi.org/10.1016/j.actamat.2015.11.042}}.
\newline\urlprefix\url{https://www.sciencedirect.com/science/article/pii/S1359645415300938}

\bibitem{MAA2018338}
R.~Maaß, P.~Derlet,
  \href{https://www.sciencedirect.com/science/article/pii/S1359645417304986}{Micro-plasticity
  and recent insights from intermittent and small-scale plasticity}, Acta
  Materialia 143 (2018) 338--363.
\newblock \href
  {http://dx.doi.org/https://doi.org/10.1016/j.actamat.2017.06.023}
  {\path{doi:https://doi.org/10.1016/j.actamat.2017.06.023}}.
\newline\urlprefix\url{https://www.sciencedirect.com/science/article/pii/S1359645417304986}

\bibitem{Fan2021}
H.~Fan, Q.~Wang, J.~A. El-Awady, D.~Raabe, M.~Zaiser,
  \href{https://doi.org/10.1038/s41467-021-21939-1}{Strain rate dependency of
  dislocation plasticity}, Nature Communications 12~(1).
\newblock \href {http://dx.doi.org/10.1038/s41467-021-21939-1}
  {\path{doi:10.1038/s41467-021-21939-1}}.
\newline\urlprefix\url{https://doi.org/10.1038/s41467-021-21939-1}

\bibitem{Chen2018}
Z.~Chen, W.~Lenthe, J.~C. Stinville, M.~Echlin, T.~M. Pollock, S.~Daly,
  \href{https://doi.org/10.1007/s11340-018-0419-y}{High-resolution deformation
  mapping across large fields of view using scanning electron microscopy and
  digital image correlation}, Experimental Mechanics 58~(9) (2018) 1407--1421.
\newblock \href {http://dx.doi.org/10.1007/s11340-018-0419-y}
  {\path{doi:10.1007/s11340-018-0419-y}}.
\newline\urlprefix\url{https://doi.org/10.1007/s11340-018-0419-y}

\bibitem{STINVILLE2020172}
J.~Stinville, P.~Callahan, M.~Charpagne, M.~Echlin, V.~Valle, T.~Pollock,
  \href{https://www.sciencedirect.com/science/article/pii/S1359645419308389}{Direct
  measurements of slip irreversibility in a nickel-based superalloy using high
  resolution digital image correlation}, Acta Materialia 186 (2020) 172--189.
\newblock \href
  {http://dx.doi.org/https://doi.org/10.1016/j.actamat.2019.12.009}
  {\path{doi:https://doi.org/10.1016/j.actamat.2019.12.009}}.
\newline\urlprefix\url{https://www.sciencedirect.com/science/article/pii/S1359645419308389}

\bibitem{MAN20101625}
J.~Man, M.~Valtr, A.~Weidner, M.~Petrenec, K.~Obrtlík, J.~Polák,
  \href{https://www.sciencedirect.com/science/article/pii/S1877705810001761}{Afm
  study of surface relief evolution in 316l steel fatigued at low and high
  temperatures}, Procedia Engineering 2~(1) (2010) 1625--1633, fatigue 2010.
\newblock \href
  {http://dx.doi.org/https://doi.org/10.1016/j.proeng.2010.03.175}
  {\path{doi:https://doi.org/10.1016/j.proeng.2010.03.175}}.
\newline\urlprefix\url{https://www.sciencedirect.com/science/article/pii/S1877705810001761}

\bibitem{Mughrabi2009}
H.~Mughrabi, \href{http://dx.doi.org/10.1007/s11661-009-9839-8}{Cyclic slip
  irreversibilities and the evolution of fatigue damage}, Metallurgical and
  Materials Transactions A 40~(6) (2009) 1257--1279.
\newblock \href {http://dx.doi.org/10.1007/s11661-009-9839-8}
  {\path{doi:10.1007/s11661-009-9839-8}}.
\newline\urlprefix\url{http://dx.doi.org/10.1007/s11661-009-9839-8}

\bibitem{Ho2015}
H.~Ho, M.~Risbet, X.~Feaugas,
  \href{http://www.sciencedirect.com/science/article/pii/S1359645414008647}{On
  the unified view of the contribution of plastic strain to cyclic crack
  initiation: Impact of the progressive transformation of shear bands to
  persistent slip bands}, Acta Materialia 85 (2015) 155 -- 167.
\newblock \href
  {http://dx.doi.org/http://doi.org/10.1016/j.actamat.2014.11.020}
  {\path{doi:http://doi.org/10.1016/j.actamat.2014.11.020}}.
\newline\urlprefix\url{http://www.sciencedirect.com/science/article/pii/S1359645414008647}

\bibitem{WELSCH2016188}
E.~Welsch, D.~Ponge, S.~{Hafez Haghighat}, S.~Sandlöbes, P.~Choi, M.~Herbig,
  S.~Zaefferer, D.~Raabe,
  \href{https://www.sciencedirect.com/science/article/pii/S1359645416304566}{Strain
  hardening by dynamic slip band refinement in a high-mn lightweight steel},
  Acta Materialia 116 (2016) 188--199.
\newblock \href
  {http://dx.doi.org/https://doi.org/10.1016/j.actamat.2016.06.037}
  {\path{doi:https://doi.org/10.1016/j.actamat.2016.06.037}}.
\newline\urlprefix\url{https://www.sciencedirect.com/science/article/pii/S1359645416304566}

\bibitem{WIECZOREK2016320}
N.~Wieczorek, G.~Laplanche, J.~Heyer, A.~Parsa, J.~Pfetzing-Micklich,
  G.~Eggeler,
  \href{https://www.sciencedirect.com/science/article/pii/S1359645416303238}{Assessment
  of strain hardening in copper single crystals using in situ sem microshear
  experiments}, Acta Materialia 113 (2016) 320--334.
\newblock \href
  {http://dx.doi.org/https://doi.org/10.1016/j.actamat.2016.04.055}
  {\path{doi:https://doi.org/10.1016/j.actamat.2016.04.055}}.
\newline\urlprefix\url{https://www.sciencedirect.com/science/article/pii/S1359645416303238}

\bibitem{Kacher2014}
J.~Kacher, B.~Eftink, B.~Cui, I.~Robertson,
  \href{https://doi.org/10.1016/j.cossms.2014.05.004}{Dislocation interactions
  with grain boundaries}, Current Opinion in Solid State and Materials Science
  18~(4) (2014) 227--243.
\newblock \href {http://dx.doi.org/10.1016/j.cossms.2014.05.004}
  {\path{doi:10.1016/j.cossms.2014.05.004}}.
\newline\urlprefix\url{https://doi.org/10.1016/j.cossms.2014.05.004}

\bibitem{Bayerschen2015}
E.~Bayerschen, A.~T. McBride, B.~D. Reddy, T.~B\"{o}hlke,
  \href{https://doi.org/10.1007/s10853-015-9553-4}{Review on slip transmission
  criteria in experiments and crystal plasticity models}, Journal of Materials
  Science 51~(5) (2015) 2243--2258.
\newblock \href {http://dx.doi.org/10.1007/s10853-015-9553-4}
  {\path{doi:10.1007/s10853-015-9553-4}}.
\newline\urlprefix\url{https://doi.org/10.1007/s10853-015-9553-4}

\bibitem{Luster1995}
J.~Luster, M.~A. Morris,
  \href{https://doi.org/10.1007/BF02670762}{Compatibility of deformation in
  two-phase ti-al alloys: Dependence on microstructure and orientation
  relationships}, Metallurgical and Materials Transactions A 26~(7) (1995)
  1745--1756.
\newblock \href {http://dx.doi.org/10.1007/BF02670762}
  {\path{doi:10.1007/BF02670762}}.
\newline\urlprefix\url{https://doi.org/10.1007/BF02670762}

\bibitem{BIELER2014212}
T.~Bieler, P.~Eisenlohr, C.~Zhang, H.~Phukan, M.~Crimp,
  \href{https://www.sciencedirect.com/science/article/pii/S1359028614000205}{Grain
  boundaries and interfaces in slip transfer}, Current Opinion in Solid State
  and Materials Science 18~(4) (2014) 212--226, slip Localization and Transfer
  in Deformation and Fatigue of Polycrystals.
\newblock \href
  {http://dx.doi.org/https://doi.org/10.1016/j.cossms.2014.05.003}
  {\path{doi:https://doi.org/10.1016/j.cossms.2014.05.003}}.
\newline\urlprefix\url{https://www.sciencedirect.com/science/article/pii/S1359028614000205}

\bibitem{HEMERY2018277}
S.~Hémery, P.~Nizou, P.~Villechaise,
  \href{https://www.sciencedirect.com/science/article/pii/S0921509317313898}{In
  situ sem investigation of slip transfer in ti-6al-4v: Effect of applied
  stress}, Materials Science and Engineering: A 709 (2018) 277--284.
\newblock \href {http://dx.doi.org/https://doi.org/10.1016/j.msea.2017.10.058}
  {\path{doi:https://doi.org/10.1016/j.msea.2017.10.058}}.
\newline\urlprefix\url{https://www.sciencedirect.com/science/article/pii/S0921509317313898}

\bibitem{SURI19991019}
S.~Suri, G.~Viswanathan, T.~Neeraj, D.-H. Hou, M.~Mills,
  \href{https://www.sciencedirect.com/science/article/pii/S1359645498003644}{Room
  temperature deformation and mechanisms of slip transmission in oriented
  single-colony crystals of an $\alpha$/$\beta$ titanium alloy}, Acta
  Materialia 47~(3) (1999) 1019--1034.
\newblock \href
  {http://dx.doi.org/https://doi.org/10.1016/S1359-6454(98)00364-4}
  {\path{doi:https://doi.org/10.1016/S1359-6454(98)00364-4}}.
\newline\urlprefix\url{https://www.sciencedirect.com/science/article/pii/S1359645498003644}

\bibitem{STINVILLE2019152}
J.~Stinville, E.~R. Yao, P.~G. Callahan, J.~Shin, F.~Wang, M.~P. Echlin, T.~M.
  Pollock, D.~S. Gianola,
  \href{https://www.sciencedirect.com/science/article/pii/S1359645419300370}{Dislocation
  dynamics in a nickel-based superalloy via in-situ transmission scanning
  electron microscopy}, Acta Materialia 168 (2019) 152--166.
\newblock \href
  {http://dx.doi.org/https://doi.org/10.1016/j.actamat.2018.12.061}
  {\path{doi:https://doi.org/10.1016/j.actamat.2018.12.061}}.
\newline\urlprefix\url{https://www.sciencedirect.com/science/article/pii/S1359645419300370}

\bibitem{Larson:2002}
B.~C. Larson, W.~Yang, G.~E. Ice, J.~D. Budai, J.~Z. Tischler,
  \href{https://doi.org/10.1038/415887a}{Three-dimensional x-ray structural
  microscopy with submicrometre resolution}, Nature 415~(6874) (2002) 887--890.
\newblock \href {http://dx.doi.org/10.1038/415887a}
  {\path{doi:10.1038/415887a}}.
\newline\urlprefix\url{https://doi.org/10.1038/415887a}

\bibitem{Poulsen2012}
H.~F. Poulsen, \href{http://dx.doi.org/10.1107/S0021889812039143}{{An
  introduction to three-dimensional X-ray diffraction microscopy}}, Journal of
  Applied Crystallography 45~(6) (2012) 1084--1097.
\newblock \href {http://dx.doi.org/10.1107/S0021889812039143}
  {\path{doi:10.1107/S0021889812039143}}.
\newline\urlprefix\url{http://dx.doi.org/10.1107/S0021889812039143}

\bibitem{Suter2006}
R.~M. Suter, D.~Hennessy, C.~Xiao, U.~Lienert,
  \href{http://link.aip.org/link/?RSI/77/123905/1}{{Forward modeling method for
  microstructure reconstruction using x-ray diffraction microscopy:
  Single-crystal verification}}, Review of Scientific Instruments 77~(12)
  (2006) 123905.
\newblock \href {http://dx.doi.org/10.1063/1.2400017}
  {\path{doi:10.1063/1.2400017}}.
\newline\urlprefix\url{http://link.aip.org/link/?RSI/77/123905/1}

\bibitem{Ludwig2009}
W.~Ludwig, P.~Reischig, A.~King, M.~Herbig, E.~M. Lauridsen, G.~Johnson, T.~J.
  Marrow, J.~Y. Buffi{\`{e}}re,
  \href{https://doi.org/10.1063/1.3100200}{Three-dimensional grain mapping by
  x-ray diffraction contrast tomography and the use of friedel pairs in
  diffraction data analysis}, Review of Scientific Instruments 80~(3) (2009)
  033905.
\newblock \href {http://dx.doi.org/10.1063/1.3100200}
  {\path{doi:10.1063/1.3100200}}.
\newline\urlprefix\url{https://doi.org/10.1063/1.3100200}

\bibitem{Bernier2020}
J.~V. Bernier, R.~M. Suter, A.~D. Rollett, J.~D. Almer,
  \href{https://doi.org/10.1146/annurev-matsci-070616-124125}{High-energy x-ray
  diffraction microscopy in materials science}, Annual Review of Materials
  Research 50~(1) (2020) 395--436.
\newblock \href {http://dx.doi.org/10.1146/annurev-matsci-070616-124125}
  {\path{doi:10.1146/annurev-matsci-070616-124125}}.
\newline\urlprefix\url{https://doi.org/10.1146/annurev-matsci-070616-124125}

\bibitem{Ludwig2007}
W.~Ludwig, E.~M. Lauridsen, S.~Schmidt, H.~F. Poulsen, J.~Baruchel,
  \href{https://doi.org/10.1107/s002188980703035x}{High-resolution
  three-dimensional mapping of individual grains in polycrystals by
  topotomography}, Journal of Applied Crystallography 40~(5) (2007) 905--911.
\newblock \href {http://dx.doi.org/10.1107/s002188980703035x}
  {\path{doi:10.1107/s002188980703035x}}.
\newline\urlprefix\url{https://doi.org/10.1107/s002188980703035x}

\bibitem{Vigano2016}
N.~Vigan{\`{o}}, A.~Tanguy, S.~Hallais, A.~Dimanov, M.~Bornert, K.~J.
  Batenburg, W.~Ludwig,
  \href{https://doi.org/10.1038/srep20618}{Three-dimensional full-field x-ray
  orientation microscopy}, Scientific Reports 6~(1).
\newblock \href {http://dx.doi.org/10.1038/srep20618}
  {\path{doi:10.1038/srep20618}}.
\newline\urlprefix\url{https://doi.org/10.1038/srep20618}

\bibitem{Nygren2020}
K.~E. Nygren, D.~C. Pagan, J.~V. Bernier, M.~P. Miller,
  \href{https://doi.org/10.1016/j.matchar.2020.110366}{An algorithm for
  resolving intragranular orientation fields using coupled far-field and
  near-field high energy x-ray diffraction microscopy}, Materials
  Characterization 165 (2020) 110366.
\newblock \href {http://dx.doi.org/10.1016/j.matchar.2020.110366}
  {\path{doi:10.1016/j.matchar.2020.110366}}.
\newline\urlprefix\url{https://doi.org/10.1016/j.matchar.2020.110366}

\bibitem{Henningsson2020}
N.~A. Henningsson, S.~A. Hall, J.~P. Wright, J.~Hektor,
  \href{https://doi.org/10.1107/S1600576720001016}{{Reconstructing
  intragranular strain fields in polycrystalline materials from scanning 3DXRD
  data}}, Journal of Applied Crystallography 53~(2).
\newblock \href {http://dx.doi.org/10.1107/S1600576720001016}
  {\path{doi:10.1107/S1600576720001016}}.
\newline\urlprefix\url{https://doi.org/10.1107/S1600576720001016}

\bibitem{Reischig_COSS_2020}
P.~Reischig, W.~Ludwig, Three-dimensional reconstruction of intragranular
  strain and orientation in polycrystals by near-field x-ray diffraction,
  Current Opinion in Solid State and Materials Science 24~(5) (2020) 100851.
\newblock \href {http://dx.doi.org/10.1016/j.cossms.2020.100851}
  {\path{doi:10.1016/j.cossms.2020.100851}}.

\bibitem{Shen2020}
Y.~F. Shen, H.~Liu, R.~M. Suter, {Voxel-based strain tensors from near-field
  High Energy Diffraction Microscopy}, Current Opinion in Solid State and
  Materials Science 24~(4) (2020) 100852.
\newblock \href {http://dx.doi.org/10.1016/j.cossms.2020.100852}
  {\path{doi:10.1016/j.cossms.2020.100852}}.

\bibitem{Miller2020}
M.~P. Miller, D.~C. Pagan, A.~J. Beaudoin, K.~E. Nygren, D.~J. Shadle,
  \href{https://doi.org/10.1007/s11661-020-05888-w}{Understanding
  micromechanical material behavior using synchrotron x-rays and in situ
  loading}, Metallurgical and Materials Transactions A 51~(9) (2020)
  4360--4376.
\newblock \href {http://dx.doi.org/10.1007/s11661-020-05888-w}
  {\path{doi:10.1007/s11661-020-05888-w}}.
\newline\urlprefix\url{https://doi.org/10.1007/s11661-020-05888-w}

\bibitem{yildirim_2020}
C.~Yildirim, P.~Cook, C.~Detlefs, H.~Simons, H.~F. Poulsen, Probing nanoscale
  structure and strain by dark-field x-ray microscopy, MRS Bulletin 45~(4)
  (2020) 277–282.
\newblock \href {http://dx.doi.org/10.1557/mrs.2020.89}
  {\path{doi:10.1557/mrs.2020.89}}.

\bibitem{Proudhon2018}
H.~Proudhon, N.~Gu{\'{e}}ninchault, S.~Forest, W.~Ludwig,
  \href{https://doi.org/10.3390/ma11102018}{Incipient bulk polycrystal
  plasticity observed by synchrotron in-situ topotomography}, Materials 11~(10)
  (2018) 2018.
\newblock \href {http://dx.doi.org/10.3390/ma11102018}
  {\path{doi:10.3390/ma11102018}}.
\newline\urlprefix\url{https://doi.org/10.3390/ma11102018}

\bibitem{Cho2020}
A.~Cho, \href{https://doi.org/10.1126/science.369.6501.234}{X-ray source gets a
  100-fold boost in brightness}, Science 369~(6501) (2020) 234--235.
\newblock \href {http://dx.doi.org/10.1126/science.369.6501.234}
  {\path{doi:10.1126/science.369.6501.234}}.
\newline\urlprefix\url{https://doi.org/10.1126/science.369.6501.234}

\bibitem{ESRF-EBS-upgrade}
{EBS} storage ring technical report,
  \url{https://www.esrf.fr/files/live/sites/www/files/about/upgrade/documentation/Design%20Report-reduced-jan19.pdf},
  accessed: 2021-08-09 (2019).

\bibitem{CHATTERJEE201635}
K.~Chatterjee, A.~Venkataraman, T.~Garbaciak, J.~Rotella, M.~Sangid,
  A.~Beaudoin, P.~Kenesei, J.-S. Park, A.~Pilchak,
  \href{http://www.sciencedirect.com/science/article/pii/S0020768316300804}{Study
  of grain-level deformation and residual stresses in ti-7al under combined
  bending and tension using high energy diffraction microscopy (hedm)},
  International Journal of Solids and Structures 94-95 (2016) 35 -- 49.
\newblock \href
  {http://dx.doi.org/https://doi.org/10.1016/j.ijsolstr.2016.05.010}
  {\path{doi:https://doi.org/10.1016/j.ijsolstr.2016.05.010}}.
\newline\urlprefix\url{http://www.sciencedirect.com/science/article/pii/S0020768316300804}

\bibitem{Lienert2009}
U.~Lienert, M.~Brandes, J.~Bernier, J.~Weiss, S.~Shastri, M.~Mills, M.~Miller,
  \href{https://doi.org/10.1016/j.msea.2009.06.047}{In situ single-grain peak
  profile measurements on ti-7al during tensile deformation}, Materials Science
  and Engineering: A 524~(1-2) (2009) 46--54.
\newblock \href {http://dx.doi.org/10.1016/j.msea.2009.06.047}
  {\path{doi:10.1016/j.msea.2009.06.047}}.
\newline\urlprefix\url{https://doi.org/10.1016/j.msea.2009.06.047}

\bibitem{Venkataraman2017}
A.~Venkataraman, P.~A. Shade, R.~Adebisi, S.~Sathish, A.~L. Pilchak, G.~B.
  Viswanathan, M.~C. Brandes, M.~J. Mills, M.~D. Sangid,
  \href{https://doi.org/10.1007/s11661-017-4024-y}{Study of structure and
  deformation pathways in ti-7al using atomistic simulations, experiments, and
  characterization}, Metallurgical and Materials Transactions A 48~(5) (2017)
  2222--2236.
\newblock \href {http://dx.doi.org/10.1007/s11661-017-4024-y}
  {\path{doi:10.1007/s11661-017-4024-y}}.
\newline\urlprefix\url{https://doi.org/10.1007/s11661-017-4024-y}

\bibitem{Kammers_2013a}
A.~Kammers, S.~Daly,
  \href{http://dx.doi.org/10.1007/s11340-013-9734-5}{Self-assembled
  nanoparticle surface patterning for improved digital image correlation in a
  scanning electron microscope}, Experimental Mechanics 53~(8) (2013)
  1333--1341.
\newblock \href {http://dx.doi.org/10.1007/s11340-013-9734-5}
  {\path{doi:10.1007/s11340-013-9734-5}}.
\newline\urlprefix\url{http://dx.doi.org/10.1007/s11340-013-9734-5}

\bibitem{Kammers_2013b}
A.~Kammers, S.~Daly, \href{http://dx.doi.org/10.1007/s11340-013-9782-x}{Digital
  image correlation under scanning electron microscopy: Methodology and
  validation}, Experimental Mechanics 53~(9) (2013) 1743--1761.
\newblock \href {http://dx.doi.org/10.1007/s11340-013-9782-x}
  {\path{doi:10.1007/s11340-013-9782-x}}.
\newline\urlprefix\url{http://dx.doi.org/10.1007/s11340-013-9782-x}

\bibitem{Stinville_2015a}
J.~Stinville, M.~Echlin, D.~Texier, F.~Bridier, P.~Bocher, T.~Pollock,
  \href{http://dx.doi.org/10.1007/s11340-015-0083-4}{Sub-grain scale digital
  image correlation by electron microscopy for polycrystalline materials during
  elastic and plastic deformation}, Experimental Mechanics (2015) 1--20\href
  {http://dx.doi.org/10.1007/s11340-015-0083-4}
  {\path{doi:10.1007/s11340-015-0083-4}}.
\newline\urlprefix\url{http://dx.doi.org/10.1007/s11340-015-0083-4}

\bibitem{LENTHE201893}
W.~C. Lenthe, J.~C. Stinville, M.~P. Echlin, Z.~Chen, S.~Daly, T.~M. Pollock,
  \href{http://www.sciencedirect.com/science/article/pii/S0304399118300305}{Advanced
  detector signal acquisition and electron beam scanning for high resolution
  sem imaging}, Ultramicroscopy 195 (2018) 93 -- 100.
\newblock \href
  {http://dx.doi.org/https://doi.org/10.1016/j.ultramic.2018.08.025}
  {\path{doi:https://doi.org/10.1016/j.ultramic.2018.08.025}}.
\newline\urlprefix\url{http://www.sciencedirect.com/science/article/pii/S0304399118300305}

\bibitem{BOURDIN2018307}
F.~Bourdin, J.~Stinville, M.~Echlin, P.~Callahan, W.~Lenthe, C.~Torbet,
  D.~Texier, F.~Bridier, J.~Cormier, P.~Villechaise, T.~Pollock, V.~Valle,
  \href{http://www.sciencedirect.com/science/article/pii/S135964541830541X}{Measurements
  of plastic localization by heaviside-digital image correlation}, Acta
  Materialia 157 (2018) 307 -- 325.
\newblock \href
  {http://dx.doi.org/https://doi.org/10.1016/j.actamat.2018.07.013}
  {\path{doi:https://doi.org/10.1016/j.actamat.2018.07.013}}.
\newline\urlprefix\url{http://www.sciencedirect.com/science/article/pii/S135964541830541X}

\bibitem{Valle2015}
V.~Valle, S.~Hedan, P.~Cosenza, A.-L. Fauchille, M.~Berdjane, {DIC Development
  for the Study of Materials Including Multiple Crossing Cracks}, Experimental
  Mechanics 55~(2) (2015) 379--391.
\newblock \href {http://dx.doi.org/10.1007/s11340-014-9948-1}
  {\path{doi:10.1007/s11340-014-9948-1}}.

\bibitem{Charpagne2019}
M.-A. Charpagne, F.~Strub, T.~M. Pollock,
  \href{https://doi.org/10.1016/j.matchar.2019.01.033}{Accurate reconstruction
  of {EBSD} datasets by a multimodal data approach using an evolutionary
  algorithm}, Materials Characterization 150 (2019) 184--198.
\newblock \href {http://dx.doi.org/10.1016/j.matchar.2019.01.033}
  {\path{doi:10.1016/j.matchar.2019.01.033}}.
\newline\urlprefix\url{https://doi.org/10.1016/j.matchar.2019.01.033}

\bibitem{Lud01a}
W.~Ludwig, P.~Cloetens, J.~H{\"{a}}rtwig, J.~Baruchel, B.~Hamelin, P.~Bastie,
  {Three-dimensional imaging of crystal defects by 'topo-tomography'}, J. Appl.
  Cryst. 34 (2001) 602--607.

\bibitem{Ludwig:2007ab}
W.~Ludwig, E.~M. Lauridsen, S.~Schmidt, H.~F. Poulsen, J.~Baruchel,
  \href{http://dx.doi.org/10.1107/S002188980703035X}{{High-resolution
  three-dimensional mapping of individual grains in polycrystals by
  topotomography}}, Journal of Applied Crystallography 40~(5) (2007) 905--911.
\newblock \href {http://dx.doi.org/10.1107/S002188980703035X}
  {\path{doi:10.1107/S002188980703035X}}.
\newline\urlprefix\url{http://dx.doi.org/10.1107/S002188980703035X}

\bibitem{dct_code}
Diffraction contrast tomography analysis code,
  \url{https://sourceforge.net/projects/dct/}, accessed: 2021-11-1 (2021).

\bibitem{Pymicro}
A.~Marano, C.~Ribart, F.~N'Guyen, H.~Proudhon, Pymicro: a data platform for
  multimodal mechanics of material microstructures, Current Opinion in Solid
  State and Materials Science.

\bibitem{Vigano_COSS_2020}
N.~Vigan\`o, W.~Ludwig, X-ray orientation microscopy using topo-tomography and
  multi-mode diffraction contrast tomography, Current Opinion in Solid State
  and Materials Science 24~(4) (2020) 100832.
\newblock \href {http://dx.doi.org/10.1016/j.cossms.2020.100832}
  {\path{doi:10.1016/j.cossms.2020.100832}}.

\bibitem{Gueninchault2016}
N.~Gueninchault, H.~Proudhon, W.~Ludwig,
  \href{https://doi.org/10.1107/s1600577516013850}{Nanox: a miniature
  mechanical stress rig designed for near-field x-ray diffraction imaging
  techniques}, Journal of Synchrotron Radiation 23~(6) (2016) 1474--1483.
\newblock \href {http://dx.doi.org/10.1107/s1600577516013850}
  {\path{doi:10.1107/s1600577516013850}}.
\newline\urlprefix\url{https://doi.org/10.1107/s1600577516013850}

\bibitem{Echlin2021}
M.~P. Echlin, M.~Kasemer, K.~Chatterjee, D.~Boyce, J.~C. Stinville, P.~G.
  Callahan, E.~Wielewski, J.-S. Park, J.~C. Williams, R.~M. Suter, T.~M.
  Pollock, M.~P. Miller, P.~R. Dawson,
  \href{https://doi.org/10.1007/s11661-021-06233-5}{Microstructure-based
  estimation of strength and ductility distributions for alpha +beta titanium
  alloys}, Metallurgical and Materials Transactions A 52~(6) (2021) 2411--2434.
\newblock \href {http://dx.doi.org/10.1007/s11661-021-06233-5}
  {\path{doi:10.1007/s11661-021-06233-5}}.
\newline\urlprefix\url{https://doi.org/10.1007/s11661-021-06233-5}

\bibitem{CHARPAGNE2021117037}
M.~Charpagne, J.~Hestroffer, A.~Polonsky, M.~Echlin, D.~Texier, V.~Valle,
  I.~Beyerlein, T.~Pollock, J.~Stinville,
  \href{https://www.sciencedirect.com/science/article/pii/S1359645421004171}{Slip
  localization in inconel 718: A three-dimensional and statistical
  perspective}, Acta Materialia 215 (2021) 117037.
\newblock \href
  {http://dx.doi.org/https://doi.org/10.1016/j.actamat.2021.117037}
  {\path{doi:https://doi.org/10.1016/j.actamat.2021.117037}}.
\newline\urlprefix\url{https://www.sciencedirect.com/science/article/pii/S1359645421004171}

\bibitem{Tanner1976}
B.~Tanner, {X-ray diffraction topography}, Pergamon Press, 1976.

\bibitem{Alcala2020}
J.~Alcal{\'a}, J.~O{\v{c}}en{\'a}{\v{s}}ek, J.~Varillas, J.~A.~El-Awady, J.~M.
  Wheeler, J.~Michler,
  \href{https://doi.org/10.1038/s41598-020-75934-5}{Statistics of dislocation
  avalanches in fcc and bcc metals: dislocation mechanisms and mean swept
  distances across microsample sizes and temperatures}, Scientific Reports
  10~(1) (2020) 19024.
\newblock \href {http://dx.doi.org/10.1038/s41598-020-75934-5}
  {\path{doi:10.1038/s41598-020-75934-5}}.
\newline\urlprefix\url{https://doi.org/10.1038/s41598-020-75934-5}

\bibitem{GUO2015229}
Y.~Guo, D.~Collins, E.~Tarleton, F.~Hofmann, J.~Tischler, W.~Liu, R.~Xu,
  A.~Wilkinson, T.~Britton,
  \href{https://www.sciencedirect.com/science/article/pii/S1359645415003626}{Measurements
  of stress fields near a grain boundary: Exploring blocked arrays of
  dislocations in 3d}, Acta Materialia 96 (2015) 229--236.
\newblock \href
  {http://dx.doi.org/https://doi.org/10.1016/j.actamat.2015.05.041}
  {\path{doi:https://doi.org/10.1016/j.actamat.2015.05.041}}.
\newline\urlprefix\url{https://www.sciencedirect.com/science/article/pii/S1359645415003626}

\bibitem{Villechaise_2012}
P.~Villechaise, J.~Cormier, T.~Billot, J.~Mendez, Mechanical behaviour and
  damage processes of {Udimet 720Li}: influence of localized plasticity at
  grain boundaries, in: 12th International Symposium on Superalloys, 2012, pp.
  15--24.

\bibitem{Polcarova2006}
M.~Polcarova, J.~Gemperlova, A.~Jacques, J.~Bradler, A.~George, {Synchrtron
  radiation topographic study of slip transfer across grain boundaries in Fe-Si
  bicrystals}, J. Phys. D: Appl. Phys. 39 (2006) 4440.

\end{thebibliography}

\newpage

\end{document}